\documentclass[conference]{IEEEtran}

\usepackage{array}

\usepackage{pifont}%
\usepackage{graphicx}
\usepackage[dvipsnames]{xcolor}
\usepackage{xspace}
\usepackage{listings}

\usepackage{cleveref}
\usepackage{xurl}

\usepackage{booktabs}
\usepackage{multirow}
\usepackage{graphicx}
\usepackage{pifont}%
\newcommand{\cmark}{\ding{51}}%
\newcommand{\xmark}{\ding{55}}%
\newcommand{\omark}{\ding{104}}

\newcommand{\codename}{{\sc Sigy}\xspace}

\newcommand{\etal}{et al.\xspace}

\newcommand{\loc}{LoC\xspace}

\newcommand{\sigfpe}{\texttt{sigfpe}\xspace}
\newcommand{\sigaction}{\texttt{rt\_sigaction}\xspace}

\newcommand{\sigusrtwo}{\texttt{sigusr2}\xspace}

\newcommand{\sigsegv}{\texttt{sigsegv}\xspace}

\newcommand{\sigusrone}{\texttt{sigusr1}\xspace}
\newcommand{\sighup}{\texttt{sighup}\xspace}

\newcommand{\raisesig}{\texttt{raise}\xspace}
\newcommand{\tkill}{\texttt{tkill}\xspace}

\newcommand{\sgxone}{SGX 1\xspace}
\newcommand{\sgxtwo}{SGX 2\xspace}

\newcommand{\ecallsighandle}{\texttt{t\_signal\_handler\_ecall}\xspace}

\newcommand{\ocallraise}{\texttt{u\_raise\_ocall}\xspace}

\newcommand{\palarithmeticexception}{\texttt{pal\_event\_arithmetic\_error}\xspace}

\newcommand{\arithmeticexception}{\texttt{ArithmeticException}\xspace}
\newcommand{\divideerror}{\texttt{DivideError}\xspace}

\newcommand{\nginx}{Nginx\xspace}

\newcommand{\nodejs}{Node.js\xspace}

\crefname{figure}{Figure}{Figure}
\crefname{section}{Section}{Section}
\crefname{appendix}{Appx.}{Appx.}
\crefname{table}{Table}{Table}
\crefname{listing}{Listing}{Listing}

\begin{document}
\title{\codename: Breaking Intel SGX Enclaves with\\  Malicious Exceptions  
\& Signals}

\author{\IEEEauthorblockN{Supraja Sridhara, Andrin Bertschi, Benedict Schl\"uter, Shweta Shinde}
\IEEEauthorblockA{ETH Z\"urich}}

\maketitle

\begin{abstract}
User programs recover from hardware exceptions and respond to signals by executing custom handlers that they register specifically for such events. 
We present \codename attack, which abuses this programming model on Intel SGX to break the confidentiality and integrity guarantees of enclaves. 
\codename uses the untrusted OS to deliver fake hardware events and injects fake signals in an enclave at any point.
Such unintended execution of benign program-defined handlers in an enclave  corrupts its state and violates execution integrity. 
7 runtimes and library OSes (OpenEnclave, Gramine, Scone, Asylo, Teaclave, Occlum, EnclaveOS) are vulnerable to \codename. 8 languages supported in Intel SGX have programming constructs that are vulnerable to \codename.
We use \codename to demonstrate 4 proof of concept exploits on webservers (Nginx, Node.js) to leak secrets and data analytics workloads in different languages (C and Java) to break execution integrity. 

\end{abstract}

\section{Introduction}

User programs rely on exceptions and signals to manage unexpected events or errors that may occur during execution.
Programming languages allow developers to express event and error-specific logic in the form of {\em handlers}, such that when the program is notified of a particular event, the handler is executed automatically.
These rich abstractions are facilitated by the cooperation of the hardware and the operating system (OS). 
When the hardware encounters runtime errors during the program execution (e.g., page faults, segmentation faults, timers, divide by zero), it notifies the OS via interrupts. The OS either handles the faults itself (e.g., load the page into memory) or forwards it to the user program's signal handler (e.g., DivideByZeroException).
The program can also request the OS for notifications about events of its interest that either emanate from the system (e.g., Ctrl+C) or other processes on the system (e.g., synchronization between parent and child processes) and register handlers that should execute on such events~\cite{posix-standard}. 
Thus, the OS not only monitors for such events on behalf of the application and notifies it, but also diverts the control of the application to the event handlers in the program. 
Both these mechanisms facilitate rich functionality in the programs, while the hardware and the OS provide efficient notification and handler invocation. 

Intel SGX provides a user-level abstraction called {\em enclaves} that protects sensitive data and code execution~\cite{sgx,costan2016intel}. 
The hardware protects enclave confidentiality and integrity even when the OS and other user processes are compromised. 
With such a strong threat model, Intel SGX limits the attack surface to the critical code running in an enclave. 
Since the enclave memory is rendered inaccessible to the OS, traditional programs that were written with the assumption of a trusted OS simply cannot execute inside enclaves (e.g., $\tt{syscall}$ instruction is illegal inside an enclave).

Due to its unique programming model, existing programs do not execute out of the box on Intel SGX~\cite{scone,edgelessrt,ego,mystikos,occlum,openenclave,enclaveos,gotee,gramine-gh,asylo,rustedp,enarx,teaclave,intel-sdk-sgx}.
As a solution, programmers can use SGX runtimes that provide a small trusted runtime that interfaces with the SGX hardware to expose a new high-level interface to the programmer. Alternatively, programmers can use a trusted library OS inside an enclave that can execute unmodified applications that were not programmed for enclaves. 
Runtimes and Library OSes for Intel SGX support exception and signal delivery to enclaves since it is a much-required feature for programs.
The hardware or the OS can inform the enclave about an exception or a signal by inducing an asynchronous exit. 
The enclave safely stores its current execution state and exits to untrusted code. 
The enclave can then be re-entered from another fixed entry point to execute corresponding pre-registered handlers for the exceptions or signals. 
During this flow, the hardware and the trusted software ensure that the OS cannot subvert the execution of the enclave---it executes the handler and then resumes the execution at the point where it was interrupted. 
This mechanism allows enclaves to handle runtime events even when the OS is untrusted. 

Heckler and WeSee introduce a new class of attacks called Ahoi attacks where an attacker uses notifications to compromise {Confidential VMs} (CVMs) enabled by Intel TDX and AMD SEV-SNP~\cite{heckler-usenix, wesee-oakland}. 
Ahoi attacks use interrupts under the control of a malicious hypervisor that can trigger interrupt handlers in the CVMs. 
These interrupt handlers alter the global execution state of the CVMs and compromise them. 
In light of these findings, we revisit Intel SGX and analyze if an attacker can use notifications to compromise the security of enclaves. 
We investigate two lines of inquiry: (i) what events can the OS fake to trigger handler execution inside the enclave? and (ii) can such handler execution bring about direct changes to the enclave's global execution state (e.g., variables)? 

In this paper, we introduce a new attack called \codename where the OS compromises the enclave execution by inducing fake events and signals to execute benign handlers registered by the enclave.
Intuitively, enclaves want to recover from a divide-by-zero and expect signals from another enclave. 
To handle such events, enclaves will register handlers that explicitly update the enclave state, say by changing the denominator to a non-zero value or invoking an event handler to respond to another enclave's request. 
If the OS convincingly tricks the enclave into falsely believing that such an event occurred, the enclave will stop its current execution and execute the handler that will explicitly update the enclave state (e.g., change a variable to a non-zero value or execute a function). 
In the least, this will result in corruption of the enclave's state resulting in a crash. If the OS injects the event at an opportune moment, it can use the effects of the handler to compromise the enclave.
We demonstrate this phenomenon by introducing a new attack called \codename, which exploits the OS's ability to fake signals to execute enclave handlers and subvert SGX guarantees.

We show that existing runtimes, library OSes, and programming language constructs are vulnerable to \codename.
We first analyze existing support to execute SGX applications: 8 runtimes (Intel SGX SDK, Open Enclave, Teaclave SGX-SDK, Asylo, Rust EDP, GoTEE, Enarx, and EGo) and 6 library OSes (Gramine, Scone, EnclaveOS, EdgelessRT, MystikOS, and Occlum). Then, we analyze the signal delivery mechanism and handlers for programs written in 9 languages (C, C++, Java, Python, Go, JavaScript, Rust, Julia and Wasm) to observe their behavior in enclaves.
We find that 3/8 runtimes (Open Enclave, Teaclave SGX-SDK, and Asylo) and 4/6 library OSes (Gramine, Scone, EnclaveOS, and Occlum) are susceptible to \codename because they do not detect the fake signals injected by the OS. 
Of the 9 languages we study, 8 (C, C++, Java, Python, Go, JavaScript, Rust, and Julia) offer language constructs for programs to register custom handlers. 
We hand-code applications in each of these languages to register handlers and execute them in vulnerable SGX runtimes and library OSes to confirm that they are indeed vulnerable to \codename. 
Next, we demonstrate that \codename breaks the confidentiality and integrity of 4 open-source applications (Nginx, Node.js, machine learning) that have been ported to Intel SGX by prior works. Our proof of concept exploits on these enclaves leak secrets and change outputs. 
Depending on the victim enclave, \codename may require injecting signals at a particular window of execution. 
We construct a proof of concept exploit against a worst-case application, a multi-layer perceptron, that requires 186 billion injections to bias the output. 
 
Our proposed software defenses serve as point-wise solutions against \codename. 
The vulnerable runtimes and library OSes have to make a design choice between either disabling functionality for security or leaving the onus on the developer to reason about the security. 
While the latter can be a pragmatic solution, new attacks like \codename serve as an example that programmers using runtimes and library OSes for lift-and-shift should not be burdened with this decision. 
We conclude that some programs simply cannot be protected without limiting functionality. 
Our conclusion encourages runtime and library OS maintainers to disable vulnerable exception and signal delivery interfaces. 
Our detailed analysis of existing enclave ecosystems spanning from runtimes, library OSes, programming language support, and existing enclave-bound applications provides an in-depth exploration to help future lift-and-shift solutions in making judicious choices. 
In summary, \codename brings attention to a new attack surface that requires a re-examination of the enclave ecosystem.

\noindent{\bf Contributions.} The paper makes three main contributions:
\begin{enumerate}
\item {\em \codename Attack.} We present a novel attack on Intel SGX where a malicious OS can send fake signals enclave and trick them into executing  enclave-registered handlers that change the enclave state. 
\item {\em Analysis.} 
Of the 14 popular frameworks for running applications in Intel SGX, 7 are vulnerable to \codename because they forward fake signals to the enclave whereas 8/9 popular languages used for enclave programming support custom handlers.
\item {\em Exploits.} 
We build exploits on 4 open-source enclave applications to demonstrate \codename. 
\end{enumerate}

\noindent{\bf Responsible Disclosure.}  
As per the responsible disclosure guidelines, we have informed all the $7$ impacted runtimes and library OSes.
We are coordinating patching with the maintainers and will update the paper once the process concludes.

\section{\codename Overview}
For functionality, applications register custom exception and signal handlers that alter the global state of the program.
To preserve this functionality on Intel SGX, runtimes and library OSes provide mechanisms to send signals to applications that execute in enclaves. 
\codename tricks the benign runtime and library OS signal handling mechanisms which results in the enclave executing the handlers. 
Therefore, the custom handlers in applications put together with the signal propagation infrastructure of runtimes and library OSes render the enclaves vulnerable to \codename. 

\noindent{\bf Threat Model.}
We trust the hardware in the Intel CPU package and assume that it is free from bugs. 
The enclaves are launched and attested according to the Intel SGX specification, and all software that executes inside the enclave is assumed to be bug-free. 
This includes the enclave application code, trusted runtime, and library OSes. 
We assume that all software that executes outside the enclave, including the OS, untrusted runtime, and other processes are untrusted and can be malicious. 

\subsection{Motivating Examples}
\label{ssec:inject-malicious-signals}
\noindent{\bf Faking hardware exceptions.}
\begin{figure}
    \centering
    \includegraphics[scale=0.55]{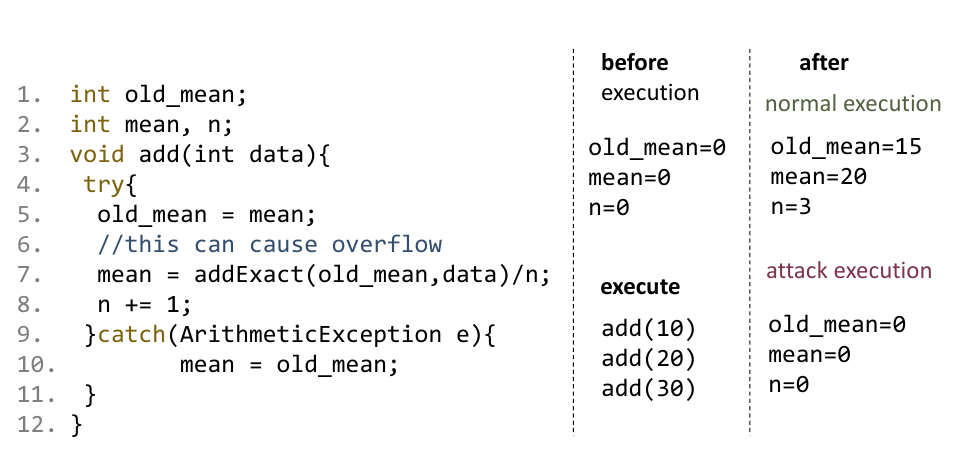}
    \caption{\codename on Java applications. Attacker injects \sigfpe $3$ times to change the execution and data integrity.}
    \label{fig:java-attack}
\end{figure}
Consider a Java function in \cref{fig:java-attack} that executes in an enclave whose \texttt{add} function (Line 3) is called $3$ times with data as $10,20,30$. The function computes and stores the new mean on every invocation.
Now, consider a malicious operating system (OS) that wants to compromise the data integrity of this enclave. 
In this example, the attacker's goal is to ensure that the application never updates the \texttt{mean}, i.e., it remains $0$ despite the $3$ invocations of \texttt{add}. 
Because this code executes in an enclave, the OS cannot change the values in memory directly.
However, observe that this enclave catches and handles \arithmeticexception{}s to deal with bad data. The handler reverts the value of \texttt{mean}, discarding the effects of the bad data.
If the OS can trigger this handler every time the \texttt{add} function is invoked, then the \texttt{mean} will not be updated. 

The OS can use \codename to achieve its goal.
Specifically, the Java runtime converts the signal for floating point exceptions (\sigfpe) that it gets from the OS to an \arithmeticexception. 
This exception is then caught and handled by the enclave.
Therefore, using \codename the OS injects \sigfpe to the Java code in the enclave when it executes Lines $4$-$9$ in~\cref{fig:java-attack}. This will result in the enclave always executing the exception handler. 
With $3$ such signal injections, the OS ensures that the \texttt{mean} does not change, thus breaking integrity. 

\noindent{\bf Faking user-defined signals.}
Several library OSes (e.g., Gramine, Scone) allow lifting and shifting unmodified applications like \nginx to execute in enclaves. 
\nginx is a web server that is highly optimized to provide maximum uptime, configured using a \texttt{config} file. 
Consider an unmodified \nginx server that executes in an enclave to serve \texttt{http} data with \texttt{config} in~\cref{lst:old-config}.
At a later point, an \nginx administrator introduces authentication tokens (JSON Web Tokens (jwt)~\cite{jwt-rfc}) in the \texttt{config} file (~\cref{lst:new-config}).
To enable the administrator to refresh the \texttt{config} file without downtime, \nginx performs the configuration refresh on receiving \sighup. 
When the administrator sends \sighup to the \nginx process, the \nginx process reads the new \texttt{config} file and starts using it. 

\begin{minipage}[t]{0.45\columnwidth}
\begin{lstlisting}[caption={Old \texttt{config}}, label={lst:old-config}]
...
server {
  listen 80;
   location /products/ {
     ...


     
  }
}
\end{lstlisting}
\end{minipage}
\begin{minipage}[t]{0.45\columnwidth}
\begin{lstlisting}[caption={New \texttt{config}}, label={lst:new-config}]
...
server {
  listen 80;
  location /products/{
  auth_jwt  "API";
  auth_jwt_type 
            encrypted
    ...
  }
}
\end{lstlisting}
\end{minipage}

Now, consider a malicious OS that aims to disable this authentication mechanism in the \nginx server. 
To do this, the OS should be able to force the server to use the old configuration file without the authentication enabled. 
Library OSes protect enclave files by encrypting them. Therefore, the malicious OS cannot directly edit the \texttt{config} file.
However, the OS can capture the old \texttt{config} file as an encrypted blob from the file-system before it is replaced by the administrator. 
Then, once the \nginx configuration upgrade is complete, and checked by the administrator, the OS writes the encrypted blob back to replace the new \texttt{config} file. Note that, this is not sufficient to trick the \nginx server into using the compromised configuration without restarting the enclave. 
The OS's goal is to ensure that when users connect to the \nginx server, they are served using the older configuration instead of the configuration that the administrator upgraded and checked, thus mounting a time-of-check time-of-use (TOCTOU) attack. 
For this, the OS uses \codename to force the \nginx server to use this compromised configuration after the administrator checks that the configuration reload was successful by injecting \sighup to the \nginx process.

\subsection{\codename Attacks on Real-world Enclaves}
\codename uses asynchronous signal injection to compromise enclaves by triggering expressive signal handlers. 
This requires a runtime or library OS that propagates hardware exceptions and signals to the enclave application. 
Further, \codename uses handlers in the applications that perform computations that alter the enclave's global state. 
The enclave handlers depend on programming language constructs used in the application (e.g., signal registration constructs, and signal handling constructs). 
Therefore, we first evaluate $14$ runtimes and library OSes and check if they propagate signals to the enclaves (\cref{sec:sdks} and~\cref{sec:liboses}).
Next, we examine $9$ programming languages and systematically analyze the constructs they provide for programs to register and execute custom signal handlers (\cref{ssec:analysis-langs}).
Finally, we use the insights from our runtime and programming language analysis to demonstrate \codename on $4$ publicly available enclave applications (\cref{sec:case-studies}). We extract secrets to break confidentiality and change programming execution to break integrity.

\section{Faking signals in SDK\MakeLowercase{s}}
\label{sec:sdks}

\noindent{\bf Background: Sending signals to threads.}
SDKs are typically used to execute a single process in an enclave. So, they may not enable mechanisms for the enclave to send signals to other processes. 
However, they do support multi-threading inside the same enclave. 
For sending signals between threads, the OS exposes the \tkill system call. 
Along similar lines, some SDKs add mechanisms to enable enclave threads to send signals to each other. 
To support sending signals to other threads, the SDKs add a new ocall interface to send a signal to the target enclave thread via the OS (Steps 1-4 in~\cref{fig:sdk-exceptions}(b)).
When the OS sends the signal to the target enclave thread, the untrusted runtime's handler propagates it to the trusted runtime using an ecall (Step 5) which in turn invokes the target thread's signal handler in the enclave.

\subsection{Intel SGX SDK}
\label{ssec:intel-sgx-sdk}
\begin{figure}
    \centering
    \includegraphics[scale=0.49]{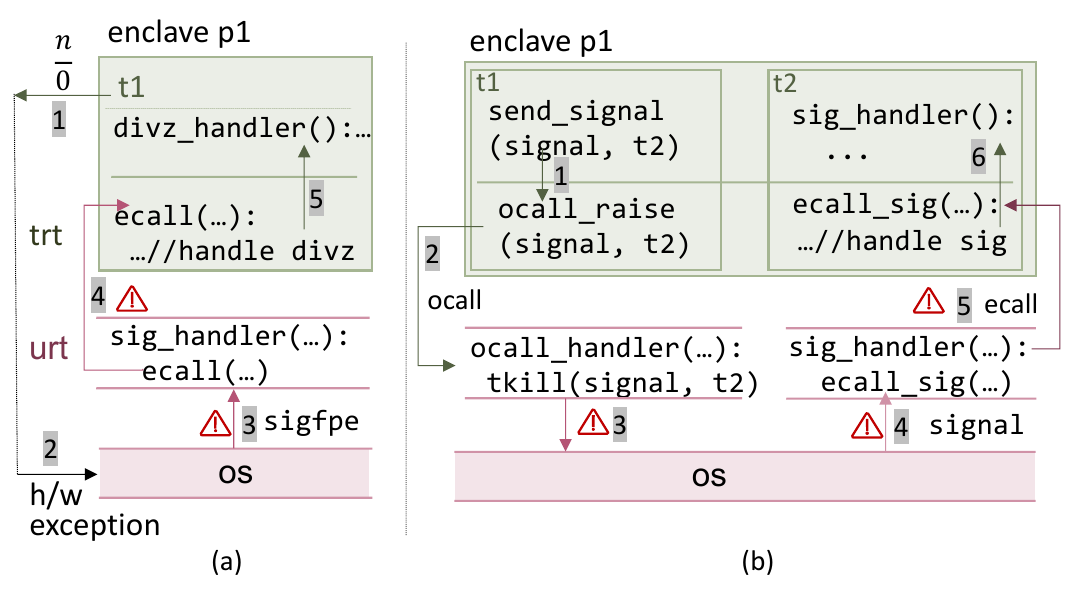}
    \caption{SDK interfaces. (a) Handling hardware exceptions (b) Handling intra-enclave signals.}
    \label{fig:sdk-exceptions}
\end{figure}
The Intel SGX SDK allows enclaves to register handlers for hardware exceptions (e.g., divide-by-zero, overflow) 
to perform custom handling when hardware exceptions occur. 
During normal operation, if the enclave executes an instruction that triggers a hardware exception (\cref{fig:sdk-exceptions}(a)), the SGX hardware triggers an asynchronous exit. 
On an asynchronous exit, the trusted hardware stores the current execution state of the enclave into the protected state save area (SSA). 
Then, it raises an exception that the untrusted OS traps on where the hardware stores information about the exit into the SSA  as shown in~\cref{lst:exit-info}. 
Crucially, in SGX 2, the hardware stores information about the exit into the SSA as shown in~\cref{lst:exit-info}. 
Note that, SGX 1 does not store this hardware information~\cite{smashex2021}. 
\begin{lstlisting}[language=C, caption={Exit information stored in the SSA.}, label=lst:exit-info]
struct _exit_info_t {
 uint32_t  vector;    // exception vector number
 uint32_t  exit_type; // HW or SW exceptions
 uint32_t  valid;     // supported/unsupported         
};

\end{lstlisting}
This includes the validity, type, and reason (i.e., exception vector) for the asynchronous exit. 
Enclave software can access this information while untrusted software (e.g., untrusted runtime, OS) cannot. 
The Intel SGX SDK's trusted runtime uses the exit information from the SSA to deduce the validity (see Line $1$ in \cref{lst:sdk-exception-handling}) and reason for any asynchronous exit (Line $4-5$).
\begin{lstlisting}[language=C, caption={Psuedocode of exception handling for Intel SGX SDK.}, label={lst:sdk-exception-handling}]
if (ssa_gpr->exit_info.valid == 1) {
  // info used to forward exception to enclave app
  info->exception_valid = ssa_gpr->exit_info.valid
  info->exception_vector = ssa_gpr->exit_info.vector;
  info->exception_type = ssa_gpr->exit_info.exit_type;
  ...
}
\end{lstlisting}
When a hardware exception causes an exit from the enclave (Step 1 in~\cref{fig:sdk-exceptions}(a)), the trusted hardware saves the exit information in the SSA and raises a hardware exception to the OS (Step 2). 
The OS converts the hardware exception to a signal.\footnote{OS does not raise a signal for page faults. It simply resumes operation in the untrusted runtime} Then, it identifies the enclave process that caused the exception and sends it a  signal (Step 3). 
This signal is caught and handled by the untrusted runtime of the Intel SGX SDK. 
The signal handler converts the signal back to the corresponding hardware exception. Then, it notifies the enclave with the hardware exception information by performing an enclave enter (Step 4).\footnote{The trusted runtime implements $2$-level exception handling for asynchronous exits. We abstract this detail in our discussion.}

\noindent{\bf \sgxone.} 
The trusted runtime uses the hardware exception vector from the untrusted runtime to call the enclave application's exception handler. 
Therefore, \sgxone enclaves that do not have hardware support to store the exit information are vulnerable to \codename. 
Specifically, the OS can arbitrarily inject a signal to the enclave to cause an asynchronous exit. 
Then, the untrusted runtime enters the enclave with the hardware exception vector corresponding to the signal. 
The exception handling in the trusted runtime executes the enclave's exception handler (Step 5) without any filtering.
Therefore, an attacker can use \codename to asynchronously send signals to the enclave and trigger its exception handlers.

\noindent{\bf \sgxtwo.} Enclaves that execute in \sgxtwo with the Intel SGX SDK are not vulnerable to \codename. 
In \sgxtwo, the trusted runtime uses the exit information from the SSA to determine the validity and exception vector when performing exception handling. 
If the OS maliciously injects signals to the enclave, it causes an invalid exit (i.e., the hardware stores 0 in Line $4$ in~\cref{lst:exit-info}). 
When the trusted runtime checks the exit information, the guard check on Line $1$ of~\cref{lst:sdk-exception-handling} will fail. 
So the trusted runtime will discard the exception and will not execute the handler in the enclave (see~\cref{fig:analysis-sdk}(a)).

\subsection{Open Enclave}
\begin{figure}[t]
    \centering
    \includegraphics[scale=0.4]{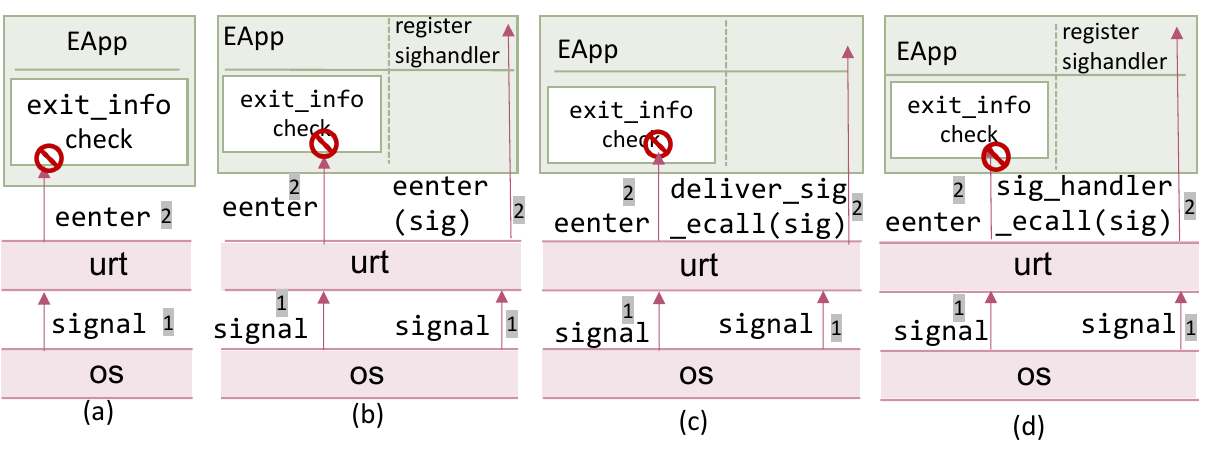}
    \caption{(a) Intel SGX SDK (b) Open enclave (c) Asylo (d) Teaclave SGX-SDK.}
    \label{fig:analysis-sdk}
\end{figure}

 Open Enclave is a widely used open-source SDK  that allows developers to write custom applications to execute in SGX enclaves.

\noindent{\bf Filtering using exit information.}
To support exception handling, Open Enclave allows applications to register handlers for hardware exceptions. 
An instruction that raises a hardware exception in the enclave causes an asynchronous exit and traps into the OS. The OS and the untrusted runtime then propagate this exception to the enclave's trusted runtime. 
Open Enclave's exception handling uses the exit information that the hardware stores in the SSA like in~\cref{lst:sdk-exception-handling}. 
If the OS attacks an enclave by maliciously injecting a signal, this will cause an asynchronous exit from the enclave. 
However, the hardware will indicate that this is an invalid exit in the exit information stored in the SSA. 
When the trusted runtime handles this exception, it will detect the attack and not execute the exception handler (LHS of~\cref{fig:analysis-sdk}(b)). 

\begin{table}[]
\caption{Library OS and Runtime analysis for \codename.  \cmark: interface supported  \xmark: interface not supported  \omark :  cannot be analyzed as they are closed source. }
\label{tab:analysis-runtimes}
\resizebox{\columnwidth}{!}{%
\begin{tabular}{c|l|cc|cc}
\hline
\multicolumn{1}{l|}{}                                                                   & Name          & \multicolumn{2}{c}{Interfaces}                                                                                                       & \multicolumn{2}{c}{Can inject with \codename}                                                                                              \\ \hline
Type                                                                                    &               & \begin{tabular}[c]{@{}c@{}}HW exception\\ interface\end{tabular} & \begin{tabular}[c]{@{}l@{}}Other signal\\ interface\end{tabular} & \begin{tabular}[c]{@{}c@{}}HW exception\\ signals\end{tabular} & \begin{tabular}[c]{@{}c@{}}Process\\ signals\end{tabular} \\ \hline
\multirow{5}{*}{SDK}                                                                    & Intel SGX SDK~\cite{intel-sdk-sgx} & \cmark                                                            & \xmark                                                           & no                                                         & no                                                   \\
                                                                                        & Open Enclave \cite{openenclave} & \cmark                                                            & \cmark                                                           & yes                                                         & yes                                                    \\
                                                                                        & Teaclave \cite{teaclave}     & \cmark                                                            & \cmark                                                           & yes                                                         & yes                                                    \\
                                                                                        & Asylo  \cite{asylo}       & \cmark                                                            & \cmark                                                           & yes                                                         & yes                                                    \\
                                                                                        & Rust EDP \cite{rustedp}     & \xmark                                                            & \xmark                                                           & no                                                         & no                                                    \\ \hline
\multirow{4}{*}{\begin{tabular}[c]{@{}c@{}}libos: 1-process\\ per enclave\end{tabular}}                  & Gramine~\cite{gramine-gh}      & \cmark                                                            & \cmark                                                           & yes                                                         & no                                                    \\
                                                                                        & Scone~\cite{scone-gh}         &  \omark                                                                 &  \omark                                                                & yes                                                         & yes                                                    \\
                                                                                        & EnclaveOS~\cite{enclaveos}     &  \omark                                                                 &  \omark                                                                & yes                                                         & yes                                                    \\
                                                                                        & EdgelessRT~\cite{edgelessrt}    & \xmark                                                            & \xmark                                                           & no                                                         & no                                                    \\ \hline
\multirow{2}{*}{\begin{tabular}[c]{@{}c@{}}libos: n-process\\ per enclave\end{tabular}} & Mystikos~\cite{mystikos}      &    \xmark                                                               &     \cmark                                                             & no                                                         & no                                                    \\
                                                                                        & Occlum~\cite{occlum}        &     \cmark                                                              &     \cmark                                                             &   yes                                                             &    yes                                                       \\ \hline
\multirow{3}{*}{\begin{tabular}[c]{@{}c@{}}language\\ runtime\end{tabular}}             & GoTEE~\cite{gotee}         &     \xmark                                                              &   \xmark                                                               & no                                                         & no                                                    \\
                                                                                        & Enarx~\cite{enarx}         &   \xmark                                                                &    \xmark                                                              & no                                                         & no                                                    \\
                                                                                        & EGo~\cite{ego}           &    \xmark                                                               &  \xmark                                                                & no                                                         & no                                                    \\ \hline
\end{tabular}%
}
\end{table}

\noindent{\bf Supporting inter-thread signals.}
The hardware exception interface allows applications to only register handlers for hardware exception events. 
Open Enclave does not allow enclaves to register handlers for other signals (e.g., \sigusrone, \sighup)
which limits the expressiveness of applications that can execute with Open Enclave. 
To improve expressiveness, Open Enclave introduces a separate mechanism to allow enclave threads to send signals to each other.  
This allows enclave threads to explicitly enable signals from the host and register signal handlers (Line $2$ and Line $3$ in~\cref{lst:oe-raise-sig}).
With this, the threads can send signals (Line $4$) to other enclave threads that are routed through the trusted runtime. 
\begin{lstlisting}[language=C, caption={Enable, register, and send a signal in Open Enclave.}, label={lst:oe-raise-sig}]
...
oe_add_vectored_exception_handler(false, sigusr_handle) 
//enable signal
oe_sgx_td_register_host_signal(thread, SIGUSR1)
//do ocall
host_send_interrupt(target_thread, SIGUSR1)
...
\end{lstlisting}

When one enclave thread wants to send a signal (e.g., \sigusrone) to the target enclave thread, it invokes the trusted runtime which performs an ocall (Step 2 in~\cref{fig:sdk-exceptions}(b)). 
In the ocall context, the untrusted runtime raises a signal to the target enclave thread using the \tkill system call. 
When the target enclave thread is resumed through \texttt{eenter} (Step 5 in~\cref{fig:sdk-exceptions}(b)), the trusted runtime calls the target enclave thread's signal handler (Step 6 in~\cref{fig:sdk-exceptions}(b)).

\noindent{\bf Attacking Open Enclave.}
When threads send signals to each other, the hardware exit information cannot be used to determine if the signals are legitimate. 
Specifically, the exit information stored in the SSA is only useful to determine the legitimacy of hardware exceptions and not explicit exits caused by sending signals. 
Therefore, the enclave cannot validate if the signal was legitimately raised by one of its threads or if it was maliciously injected by untrusted software. 
We can use the signal injection interface to compromise the security of Open Enclave using \codename. 
Concretely, untrusted software like the OS can directly send signals to enclave threads.
This causes the untrusted runtime to enter the enclave with the signal. 
Because the trusted runtime does not validate the source of this signal, it will invoke the enclaves signal handler (RHS of~\cref{fig:analysis-sdk}). 
Therefore, an attacker can arbitrarily execute the enclave's signal handler using \codename. 
\subsection{Teaclave SGX-SDK}

Teaclave SGX-SDK (Teaclave for short) enables developers to write programs in Rust and execute them in enclaves. 
It is based on Intel SGX SDK and implements different Rust libraries to ease enclave application development (i.e., standard library functionality). 

\noindent{\bf Hardware exceptions.} 
For hardware exceptions, it relies on the Intel SGX SDK's exception handling mechanism. 
The trusted runtime of Intel SGX SDK detects and discards all maliciously injected exceptions by checking the exit information in the SSA as discussed in~\cref{ssec:intel-sgx-sdk}.
Hence, this interface cannot be used to compromise Teaclave enclaves using \codename. 

\begin{figure}[t]
    \centering
    \includegraphics[scale=0.5]{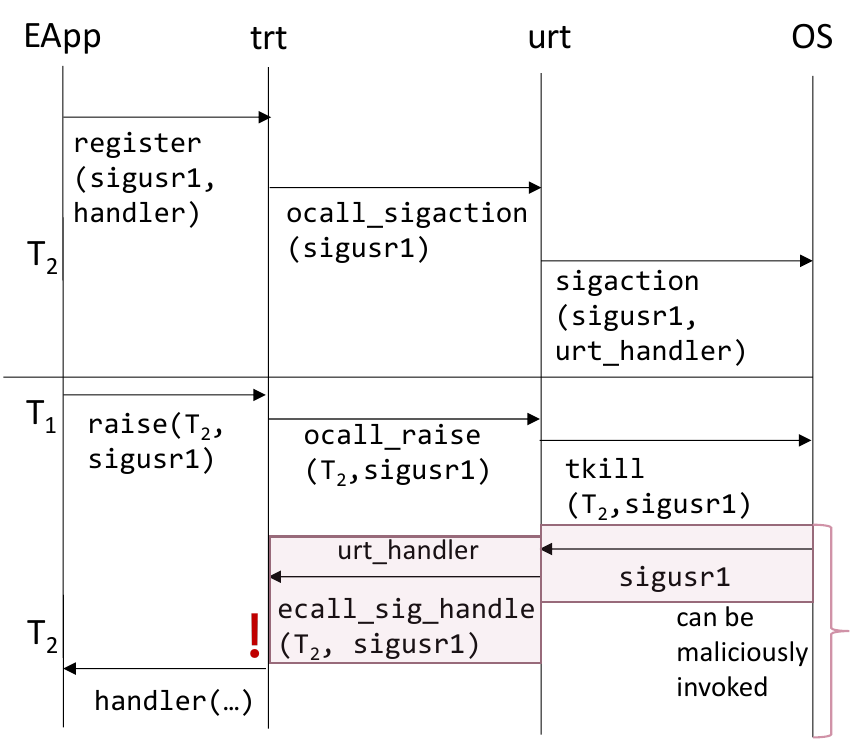}
    \caption{$T_1$ and $T_2$ are threads of the same enclave process. Black: Normal operation with ocall and ecall interfaces in Teaclave. Pink: Interfaces that can be maliciously invoked for \codename.}
    \label{fig:teaclave-sdk}
\end{figure}

\noindent{\bf Signal support.} 
Next, we analyze Teaclave's support that enables enclave threads to send signals to each other. 
Teaclave includes wrappers for Rust signal libraries. 
This allows enclave threads to register handlers and send signals to each other.
To enable this functionality, Teaclave introduces a new ocall that its Rust library wrapper invokes as shown in~\cref{fig:teaclave-sdk}. 
The ocall function in Teaclave's untrusted runtime performs \texttt{sigaction} and \raisesig libc calls 
Teaclave also adds a public ecall (\ecallsighandle) to be invoked by the untrusted runtime to forward signals from one enclave thread to another.  

During normal operation, an enclave thread can raise a signal to another thread using Teaclave's signal library. 
This is translated into an ocall (\ocallraise) by the library which transfers control to Teaclave's untrusted runtime. 
The untrusted runtime sends a signal to the target thread using \tkill. 
When the OS sends the signal to the target thread, Teaclave's untrusted runtime invokes the ecall (\ecallsighandle) to handle the signal in the enclave. 
The ecall implementation in Teaclave's trusted runtime then triggers the signal handler registered in the target thread. 

\noindent{\bf Attacking Teaclave.}
\codename needs the ability to arbitrarily inject signals that trigger the signal handler in the enclave.  
To gain this capability, \codename can abuse the \ecallsighandle interface. 
If an enclave registers signal handlers, \codename can arbitrarily inject signals into the untrusted runtime and trigger this ecall. 
Alternatively, because the untrusted runtime is attacker-controlled, \codename could directly invoke this ecall without the need for signal injection. 
In both cases, the enclave application will always execute the signal handler (\cref{fig:teaclave-sdk}). Therefore, an attacker can use \codename to trigger computation in the enclave.
Note that the trusted runtime does not have any mechanism to distinguish legitimate signals (e.g., from one enclave thread to another) from those that are maliciously injected by the OS.
Further, the \ecallsighandle is a public root ecall (i.e., invoked from any untrusted software) for functionality such that it can be invoked by the untrusted runtime to forward signals to the enclave. 
Therefore, Teaclave's signal handling design makes it vulnerable to \codename (RHS in~\cref{fig:analysis-sdk}(d)).

\subsection{Asylo}
Asylo is an open-source framework that provides a POSIX interface to enable enclave application development. 
It implements wrappers for POSIX functions that invoke ocalls to interact with the untrusted OS.
Asylo uses the Intel SGX SDK and preserves the hardware exception handling interface from the SGX SDK. 
This interface checks the exit information in the SSA and discards maliciously injected exceptions. 
So, this interface is not vulnerable to \codename. 

\noindent{\bf Signal support.} Asylo introduces a new signal handling mechanism to allow enclave applications to register and handle signals. 
In Asylo, this signal handling interface can be used by the enclave application to register all POSIX signals. 
To support signals, Asylo introduces a new ocall (\texttt{ocall\_enc\_untrusted\_register\_signal \\ \_handler}) to register a signal handler and an ecall (\texttt{ecall\_deliver\_signal}) to propagate the signal from the OS to the enclave. 
When the OS sends a signal, the untrusted runtime invokes the ecall and transfers the execution to the enclave. 
The enclaves trusted runtime uses the signal from the ecall's parameters, looks up the corresponding handler that was registered, and invokes it.  

\noindent{\bf Attacking Asylo.}
The ecall used to deliver the signal to the enclave is a public root ecall. 
Therefore, \codename can use this interface to attack enclaves in Asylo. 
Concretely, when the OS sends a signal to the enclave, the untrusted runtime invokes the ecall to enter the enclave's trusted runtime (RHS~\cref{fig:analysis-sdk}(c))
The trusted runtime executes the enclave's signal handler without any additional checks. 

In summary, our analysis shows that all SDKs use the exit information that the hardware stores to handle hardware exceptions. 
However, they introduce new mechanisms to support signal handling between threads in the enclaves which render $3$ out of the $4$ SDKs vulnerable to \codename.

\section{Faking signals in Library OS\MakeLowercase{es}}
\label{sec:liboses}
\begin{figure}
    \centering 
    \includegraphics[scale=0.7]{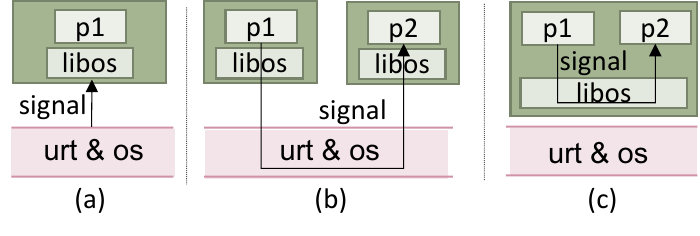}
    \caption{Signal propagation with library OSes. (a) OS or untrusted runtime sends signal to enclave process. (b) One enclave process sends signal to another enclave process through the untrusted runtime and OS. (c) LibOS creates a process abstraction such that 2 processes run in the same enclave. These processes can send signals to each other via the LibOS. }
    \label{fig:libos-signals}
\end{figure}

\noindent{\bf Background.}
Library OSes support rich exception handling and signal interfaces for enclave applications. 
Unlike the SDKs, they also implement mechanisms to execute multi-process applications by adding support for calls like \texttt{fork}, \texttt{vfork}, and \texttt{execv}. 
In some library OSes (e.g., Gramine, Scone) calls to these functions spawn new enclave processes~\cref{fig:libos-signals}(b). 
Like the OS, the library OSes also add support to send signals from one enclave process to another.
For this, they route the signals through the untrusted operating system and runtime (\cref{fig:libos-signals}(b)). 
Other library OSes support multi-processing, by implementing process abstractions inside the enclave (e.g., Occlum). 
Such library OSes route inter-process signals inside the enclave itself and do not have to exit or use the untrusted runtime and OS to propagate signals (\cref{fig:libos-signals}(c)).

\subsection{Gramine}
\label{ssec:gramine}
\begin{figure}[t]
    \centering
    \includegraphics[scale=0.55]{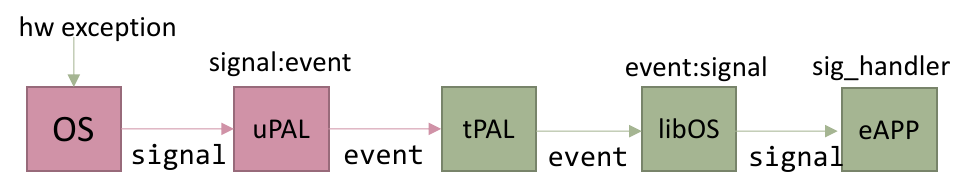}
    \caption{Hardware exception and signal handling in Gramine.}
    \label{fig:gramine}
\end{figure}
Gramine is an open-source library OS that enables executing unmodified multi-process applications in SGX enclaves. It has $3$ parts---a library OS, a Platform Adaptation Layer (PAL), and a patched C standard library.
For simplicity, we refer to the part of the PAL that runs outside the enclave as untrusted PAL (uPAL) and the part that executes inside the enclave as (tPAL). 
\cref{fig:gramine} shows exception and signal handling in Gramine.

\noindent{\bf Gramine's signal support.}
Crucially for \codename, Gramine's C standard library allows support for signal handling in enclave applications. 
Specifically, the wrappers for C's \texttt{sigaction} calls allow the application to register handlers for signals. 
When the enclave triggers a hardware exception, the untrusted OS traps on it and raises a signal that is caught by the uPAL. 
The uPAL's signal handling function (\texttt{handle\_sync\_signal}) converts the signal into a corresponding Gramine-specific PAL event. 
For example, the uPAL maps \sigfpe to \palarithmeticexception. 
Then, it invokes tPAL using the \texttt{sgx\_raise} (\cref{fig:gramine}) who forwards the event to the library OS. The library OS converts the event to a signal, creates the signal struct (\texttt{siginfo\_t}) like in Linux, and raises the signal to the enclave application. 
This finally results in the enclave executing its signal handler. 

\noindent{\bf Attacking Gramine.} The trusted PAL does not check if the event was raised because of a real hardware exception in the enclave or by a signal injected by the untrusted OS, thus making Gramine vulnerable to \codename. 
Specifically, the untrusted OS can inject a signal arbitrarily to a Gramine application (see~\cref{fig:libos-signals}(a)). This signal is converted to an event in the uPAL and forwarded to the tPAL which eventually executes the enclave application's handler. 

Although Gramine supports executing multi-process applications, it doesn't explicitly enable support for enclave processes to send signals to each other. 
Instead, its signal handling mechanism has limited expressiveness and is built to only support hardware exceptions (as shown in~\cref{tab:analysis-runtimes}). 
Therefore, with \codename we can only inject signals that map to hardware exceptions to compromise enclaves that execute with Gramine (c.f.~\cref{ssec:analysis-runtimes}).

\begin{table*}
\caption{Signal support in SDKs \& Library OSes. \cmark executes signal handler, \xmark \ crashes the enclave, \omark \ no observable behavior. (c.f. ~\cref{tab:signum-name} for mappings from signal number to signal name).}
\label{tab:analysis-vuln-runtimes}
\centering
\resizebox{2\columnwidth}{!}{%
\begin{tabular}{llllllllllllllllllllllllllllllll} 
\toprule
Runtime/LibOS         & 1 & 2 & 3 & 4* & 5* & 6 & 7* & 8* & 9 & 10 & 11* & 12 & 13 & 14 & 15 & 16 & 17 & 18 & 19 & 20 & 21 & 22 & 23 & 24 & 25 & 26 & 27 & 28 & 29 & 30 & 31*  \\
\hline
Open Enclave  &  \cmark & \xmark  &  \xmark & \xmark  & \xmark  & \cmark  & \xmark  & \xmark  & \xmark &  \cmark  & \xmark   &  \cmark  &  \cmark  &  \cmark  & \xmark   &  \xmark  & \omark    & \omark   &   \xmark  & \omark   & \omark   & \omark   & \omark   &  \xmark  & \xmark   &  \xmark  &  \xmark  &  \omark  & \omark   &  \xmark  &  \xmark   \\
Teaclave     & \cmark  & \cmark  & \cmark  & \xmark  & \xmark  & \cmark  & \xmark & \xmark  & \xmark & \cmark  & \xmark   & \cmark   & \cmark   & \cmark   & \cmark   & \cmark   &  \cmark  &  \cmark  &   \xmark  &  \cmark  &  \cmark  & \cmark   &  \cmark  & \cmark   & \cmark   &  \cmark  & \cmark   & \cmark   &  \cmark  & \cmark   & \cmark    \\
Asylo        & \cmark  & \cmark  & \cmark  & \omark  & \cmark  & \cmark  & \omark  & \cmark & \xmark & \cmark   & \cmark   & \cmark   & \cmark   & \cmark   &  \cmark  &  \cmark  & \omark   &  \cmark  &   \xmark  &  \cmark  & \cmark   &  \cmark  & \cmark   &  \cmark  &  \cmark  &  \cmark  & \cmark   &  \cmark  &  \omark  & \cmark   &   \cmark  \\
Gramine      &  \xmark & \xmark  & \xmark  & \cmark  & \xmark  & \xmark  & \cmark  & \cmark & \xmark & \xmark   & \cmark   &  \omark  &  \omark  & \xmark   & \omark   & \xmark   & \omark   &  \omark  &   \xmark  & \omark   &  \omark  & \omark   &  \omark  & \xmark   & \xmark   &  \xmark  &  \xmark  & \omark   & \xmark   &  \xmark  &   \xmark  \\
Scone        & \cmark  & \cmark  &  \cmark & \cmark  & \cmark  & \cmark  & \cmark  & \cmark & \xmark  & \cmark   & \cmark   & \cmark   & \cmark   & \cmark   & \cmark   & \cmark   & \cmark   & \cmark   &   \xmark  & \cmark   &  \cmark  & \cmark   & \cmark   & \cmark   &  \cmark  & \cmark   & \cmark   & \cmark   & \cmark   &  \cmark  & \cmark    \\
EnclaveOS    &  \cmark & \cmark  & \cmark  & \cmark  & \cmark  & \cmark  & \cmark  & \cmark & \xmark  &  \cmark  &  \cmark  &  \xmark  &  \cmark  &  \cmark  &  \cmark  &  \cmark  &  \cmark  & \cmark  &  \xmark  &  \cmark  &  \cmark  &  \cmark  &  \cmark  &  \cmark  &  \cmark  &  \cmark  &  \cmark  &  \cmark  &  \cmark  &  \cmark  &  \cmark  \\
Occlum       & \cmark  & \cmark  & \cmark  & \cmark  & \cmark  & \cmark  & \cmark  & \cmark & \xmark & \cmark   & \cmark   & \cmark   & \cmark  & \cmark   & \cmark   & \cmark   & \cmark   & \cmark   &   \xmark  &  \cmark  & \cmark   & \cmark   & \cmark   & \cmark   & \cmark   & \cmark   & \cmark   & \cmark   &  \cmark  &  \cmark  &  \cmark   \\
\bottomrule
\end{tabular}
}
\end{table*}

\subsection{Scone and EnclaveOS}
They are closed-source library OSes that support executing unmodified applications in SGX based on Intel SGX SDK. 
As we don't have the source code, we perform a black-box analysis of these library OSes. 
Specifically, we write a program that registers handlers for all signals (c.f.~\cref{ssec:analysis-runtimes}) and observe which handlers are executed. 
For both Scone~\cite{scone} and EnclaveOS~\cite{enclaveos}, our analysis shows that we can arbitrarily inject most signals from the untrusted OS to the enclaves (c.f.~\cref{ssec:analysis-runtimes}). This successfully triggers signal handlers in the enclave making them vulnerable to \codename. 
Scone forwards all signals from the OS to the enclaves.
For EnclaveOS, our analysis shows that \sigusrtwo is reserved for library OS-specific operations, and enclave applications cannot register signal handlers for these. Besides this, we observe that all other signals from the OS execute enclave handlers. 
Because these are closed-source we cannot comprehensively analyze their behavior. 
We suspect that they allow all signals because they introduce a new ecall which enables the OS to send signals to the enclaves.

\subsection{Occlum}  Occlum is an open-source library OS that enables multi-process applications by executing them in a single enclave~\cite{occlum}.  As shown in~\cref{fig:libos-signals}(c), Occlum operates under the threat model of untrusted co-resident processes inside an enclave~\cite{occlum-paper}.
It performs inter-process isolation using Intel MPX for software-fault isolation.
Crucially, it allows application processes in the enclave to send signals to each other. 
In this threat model that assumes untrusted processes inside the enclave, \codename can inject signals from one process to another. 
For Occlum, we do not need the untrusted OS to inject signals into the enclave. 
Instead, we can use malicious attacker-controlled processes to send signals to the victim process in the same enclave. 
The library OS in Occlum does not filter such injections and simply forwards the signals from the attacker-controlled process to the victim process. 
For completeness, we checked if the OS can inject signals into the enclave to trigger signal handlers.
We report that while we can inject signals, these signals do not result in the enclave executing its signal handlers. 
Instead, the library OS invokes the kernel's default signal handling which always crashes the process. 

\subsection{Other Runtimes} 
\noindent{\bf Runtimes without signal support.} 
Of the runtimes and library OSes that we surveyed, the language runtimes GoTEE, EGo, and Enarx are not vulnerable to \codename because they do not support signal handling~\cite{ego, gotee, enarx}. 
Similarly, EdgelessRT (a library OS) and RustEDP (an SDK) also do not support signal handling in enclaves~\cite{edgelessrt, rustedp}. 
Therefore, these also cannot be attacked using \codename. 

\noindent{\bf Runtimes with limited signal support.} MystikOS is a library OS that allows multiple application processes to execute in a single enclave process~\cite{mystikos}. 
Similar to Occlum, MystikOS implements mechanisms to allow application processes inside the enclave to send signals to each other. 
All signals are routed through the library OS and do not leave the enclave. 
Therefore, it doesn't expose interfaces for the untrusted software OS to inject signals. 
Unlike Occlum, it assumes that processes in the enclave mutually trust each other.
So, while processes can send signals to each other, this cannot be abused by an attacker to compromise the enclave's execution.
Therefore, MystikOS is not vulnerable to \codename.

\section{Which signals can we inject?}
\label{sec:analysis}
In~\cref{sec:sdks} and~\cref{sec:liboses}, we discussed SDKs and library OSes that are vulnerable to \codename. 
Next, we systematically analyze each of the vulnerable SDKs and library OSes to determine which signal handlers are potentially of interest to the attacker. 
Then, we examine programming language support for signals to determine if applications written in them might be vulnerable to \codename. 
\subsection{SDK and Library OS}
\label{ssec:analysis-runtimes}

\begin{table}[]
\caption{Signal number to signal name mappings. * Signals from hardware exceptions}
\label{tab:signum-name}
\centering
\resizebox{0.7\columnwidth}{!}{%
\begin{tabular}{@{}ll|ll@{}}
\toprule
\# & Name             & \# & Name           \\ \midrule
1        & SIGHUP           & 2        & SIGINT         \\
3        & SIGQUIT          & 4*        & SIGILL         \\
5*        & SIGTRAP          & 6        & SIGABRT/SIGIOT \\
7*        & SIGBUS           & 8*        & SIGFPE         \\
9        & SIGKILL          & 10       & SIGUSR1        \\
11*       & SIGSEGV          & 12       & SIGUSR2        \\
13       & SIGPIPE          & 14       & SIGALRM        \\
15       & SIGTERM          & 16       & SIGSTKFLT      \\
17       & SIGCHLD          & 18       & SIGCONT        \\
19       & SIGSTOP          & 20       & SIGTSTP        \\
21       & SIGTTIN          & 22       & SIGTTOU        \\
23       & SIGURG           & 24       & SIGXCPU        \\
25       & SIGXFSZ          & 26       & SIGVTALRM      \\
27       & SIGPROF          & 28       & SIGWINCH       \\
29       & SIGIO            & 30       & SIGPWR         \\
31*       & SIGSYS/SIGUNUSED &          &                \\ \bottomrule
\end{tabular}%
 }
\end{table}

To determine which signals are of interest to \codename in vulnerable SDKs and library OSes, we write test applications that register handlers for all signals in the range $1$-$31$.
We execute our test applications in each of the $7$ vulnerable SDKs and library OSes. 
Then, for each run of the test application we inject a signal from the OS and 
check if the application: (a) executes the registered handler, (b) crashes, or (c) has no effect as shown in~\cref{tab:analysis-vuln-runtimes}.

\begin{table*}
\caption{Signal support in Programming languages. \cmark executes signal handler,  \xmark \ crashes the program,  \omark \ no observable behavior.}
\label{tab:lang-extended-analysis}
\centering
\resizebox{1.9\columnwidth}{!}{%
\begin{tabular}{lccccccccccccccccccccccccccccccc} 
\toprule
Language         & 1 & 2 & 3 & 4 & 5 & 6 & 7 & 8 & 9 & 10 & 11 & 12 & 13 & 14 & 15 & 16 & 17 & 18 & 19 & 20 & 21 & 22 & 23 & 24 & 25 & 26 & 27 & 28 & 29 & 30 & 31  \\ \midrule
C & \cmark  & \cmark  & \cmark  & \cmark  & \cmark  & \cmark  & \cmark  & \cmark & \xmark  & \cmark  & \cmark & \cmark  & \cmark  & \cmark   & \cmark   & \cmark   & \cmark   & \cmark  & \xmark & \cmark  & \cmark   & \cmark   & \cmark   &  \cmark  & \cmark   &  \cmark  &  \cmark  &  \cmark  &  \cmark  & \cmark   &  \cmark   \\
C++ & \cmark  & \cmark  & \cmark  & \cmark  & \cmark  & \cmark  & \cmark  & \cmark & \xmark & \cmark  & \cmark  & \cmark  & \cmark  & \cmark   & \cmark   & \cmark   & \cmark   & \cmark & \xmark & \cmark   & \cmark   & \cmark   & \cmark   &  \cmark  & \cmark   &  \cmark  &  \cmark  &  \cmark  &  \cmark  & \cmark   &  \cmark   \\
Java  & \xmark   & \xmark  & \xmark  & \xmark  & \xmark  & \xmark  &  \xmark & \cmark & \xmark &  \xmark  & \xmark  & \xmark  & \xmark  & \xmark & \xmark &  \xmark  & \xmark   & \xmark  & \xmark  & \xmark  & \xmark  & \xmark   & \xmark   & \xmark   & \xmark   & \xmark   & \xmark  & \xmark   & \xmark   & \xmark   & \xmark    \\
Python & \cmark  & \cmark  & \cmark  & \cmark  & \cmark  & \cmark  & \cmark  & \cmark & \xmark &  \cmark  & \cmark & \cmark  & \cmark  & \cmark   & \cmark   & \cmark   & \cmark   & \cmark  & \xmark & \cmark   & \cmark   & \cmark   & \cmark   &  \cmark  & \cmark   &  \cmark  &  \cmark  &  \cmark  &  \cmark  & \cmark   &  \cmark   \\
Go & \cmark  & \cmark  & \cmark  & \cmark  & \cmark  & \cmark  & \cmark  & \cmark & \xmark & \cmark  & \cmark & \cmark  & \cmark  & \cmark   & \cmark   & \cmark   & \cmark   & \cmark  & \xmark & \cmark   & \cmark   & \cmark   & \cmark   &  \cmark  & \cmark   &  \cmark  &  \omark  &  \cmark  &  \cmark  & \cmark   &  \cmark   \\
JS & \cmark  & \cmark  & \cmark  & \cmark  & \cmark  & \cmark  &  \cmark  & \cmark & \xmark & \cmark  & \cmark & \cmark  & \cmark  & \cmark   & \cmark   & \cmark   & \cmark   & \cmark & \xmark  & \cmark   & \cmark   & \cmark   & \cmark   &  \cmark  & \cmark   &  \cmark  &  \cmark  &  \cmark  &  \cmark  & \cmark   &  \cmark   \\
Rust & \cmark  & \cmark  & \cmark  & \cmark  & \cmark  & \cmark  & \cmark  &  \cmark & \xmark & \cmark  & \cmark & \cmark  & \cmark  & \cmark   & \cmark   & \cmark   & \cmark   & \cmark  & \xmark & \cmark   & \cmark   & \cmark   & \cmark   &  \cmark  & \cmark   &  \cmark  &  \cmark  &  \cmark  &  \cmark  & \cmark   &  \cmark   \\
Wasm & \xmark  & \xmark  & \xmark  & \xmark  & \xmark  & \xmark  & \xmark  & \xmark & \xmark & \xmark  & \xmark & \xmark & \xmark & \xmark & \xmark  & \xmark  & \xmark  &  \xmark & \xmark & \xmark &  \xmark  & \xmark  &  \xmark  & \xmark  & \xmark & \xmark &  \xmark  & \xmark   & \xmark   &  \xmark  & \xmark    \\
Julia & \xmark  & \cmark  & \xmark  & \xmark  & \omark  & \xmark  & \xmark  & \cmark & \xmark & \cmark  & \xmark & \omark & \omark & \xmark & \xmark  & \xmark  & \omark  &  \omark & \xmark & \cmark &  \cmark  & \cmark  &  \cmark  & \xmark  & \xmark & \xmark &  \xmark  & \omark   & \xmark   &  \xmark  & \xmark    \\
\bottomrule
\end{tabular}
}
\end{table*}

\begin{table*}
\caption{Signal handlers registered by default in Interpreters/Compilers. \cmark handler registered, \xmark no handler registered.}
\label{tab:analysis-lang-sigs-default}
\centering
\resizebox{2\columnwidth}{!}{%
\begin{tabular}{llllllllllllllllllllllllllllllll} 
\toprule
Language         & 1 & 2 & 3 & 4 & 5 & 6 & 7 & 8 & 9 & 10 & 11 & 12 & 13 & 14 & 15 & 16 & 17 & 18 & 19 & 20 & 21 & 22 & 23 & 24 & 25 & 26 & 27 & 28 & 29 & 30 & 31  \\
\midrule
C & \xmark  & \xmark  & \xmark  & \xmark  & \xmark  & \xmark  & \xmark  & \xmark & \xmark  & \xmark  & \xmark & \xmark  & \xmark  & \xmark   & \xmark   & \xmark   & \xmark   & \xmark & \xmark & \xmark   & \xmark   & \xmark   & \xmark   &  \xmark  & \xmark   &  \xmark  &  \xmark  &  \xmark  &  \xmark  & \xmark   &  \xmark   \\
C ++ & \xmark  & \xmark  & \xmark  & \xmark  & \xmark  & \xmark  & \xmark  & \xmark & \xmark  & \xmark  & \xmark & \xmark  & \xmark  & \xmark   & \xmark   & \xmark   & \xmark   & \xmark & \xmark  & \xmark   & \xmark   & \xmark   & \xmark   &  \xmark  & \xmark   &  \xmark  &  \xmark  &  \xmark  &  \xmark  & \xmark   &  \xmark   \\
Java  & \cmark  & \cmark  & \cmark  & \cmark  & \xmark  & \xmark  & \cmark  & \cmark & \xmark  & \xmark  & \cmark  & \cmark  & \cmark   & \xmark   & \cmark   & \xmark   & \xmark   & \xmark & \xmark & \xmark   & \xmark   & \xmark & \xmark   &  \xmark  & \cmark   &  \xmark  &  \xmark  &  \xmark  &  \xmark  & \xmark   &  \xmark   \\ 
Python & \xmark  & \cmark  & \xmark  & \xmark  & \xmark  & \xmark  & \xmark  & \xmark & \xmark  & \xmark  & \xmark & \xmark  & \xmark  & \xmark   & \xmark   & \xmark   & \xmark   & \xmark  & \xmark & \xmark   & \xmark   & \xmark   & \xmark   &  \xmark  & \xmark   &  \xmark  &  \xmark  &  \xmark  &  \xmark  & \xmark   &  \xmark   \\ 
Go & \cmark  & \cmark  & \cmark  & \cmark  & \cmark  & \cmark  & \cmark  & \cmark & \xmark & \cmark  & \cmark & \cmark  & \cmark  & \cmark   & \cmark   & \cmark   & \cmark   & \xmark  & \xmark & \xmark   & \xmark   & \xmark   & \cmark   &  \cmark  & \cmark   &  \cmark  &  \cmark  &  \cmark  &  \cmark  & \cmark   &  \cmark   \\
JS & \xmark  & \cmark  & \xmark  & \xmark  & \xmark  & \xmark  & \xmark  & \xmark & \xmark & \cmark  & \cmark & \xmark  & \xmark  & \xmark   & \cmark   & \xmark   & \xmark   & \xmark  & \xmark & \xmark   & \xmark   & \xmark   & \xmark   &  \xmark  & \xmark   &  \xmark  &  \xmark  &  \cmark  &  \xmark  & \xmark   &  \xmark   \\ 
Rust& \xmark  & \xmark  & \xmark  & \xmark  & \xmark  & \xmark  & \cmark  & \xmark & \xmark & \xmark  & \cmark & \xmark  & \xmark  & \xmark   & \xmark   & \xmark   & \xmark   & \xmark & \xmark & \xmark   & \xmark   & \xmark   & \xmark   &  \xmark  & \xmark   &  \xmark  &  \xmark  &  \xmark  &  \xmark  & \xmark   &  \xmark   \\ 
Wasm & \xmark  & \xmark  & \xmark  & \cmark  & \xmark  & \xmark  & \cmark  & \cmark & \xmark  & \xmark  & \cmark & \xmark  & \xmark  & \xmark   & \xmark   & \xmark   & \xmark   & \xmark & \xmark  & \xmark   & \xmark   & \xmark   & \xmark   &  \xmark  & \xmark   &  \xmark  &  \xmark  &  \xmark  &  \xmark  & \xmark   &  \xmark   \\ 
Julia & \xmark  & \cmark  & \xmark  & \cmark  & \xmark  & \cmark  & \cmark  & \cmark & \xmark & \cmark  & \cmark & \cmark & \xmark & \xmark & \xmark  & \xmark  & \xmark  &  \xmark & \xmark & \xmark &  \xmark  & \xmark  &  \xmark  & \xmark  & \xmark & \xmark &  \xmark  & \xmark   & \xmark   &  \xmark  & \cmark    \\
\bottomrule
\end{tabular}
}
\end{table*}

\noindent{\bf Our findings.}
Our experiments show that when our test applications invoke the \sigaction system call and try to register handlers for \texttt{sigkill} and \texttt{sigstp}, the system call fails and the application always crashes for all SDKs and library OSes. 
This is because these signals are reserved by the operating system for process management (e.g., \texttt{sigkill} is used to force kill the process). 
Occlum and Scone allow handlers for all other signals except \texttt{sigkill} and \texttt{sigstp} to be registered and we report that their signal handlers are executed. 
EnclaveOS reserves \sigusrtwo for library OS operations but forwards all other signals except \texttt{sigkill} and \texttt{sigstp} to the application and executes its registered handlers. 
Notably, Teaclave does not allow injecting the signals that it expects to get from hardware exceptions (\texttt{sigill}, \texttt{sigtrap}, \texttt{sigbus}, \texttt{sigfpe}, and \texttt{sigsegv}) through the signal interface used for inter-thread communication. 
In contrast, Gramine only allows applications to use signals that map to a limited set of hardware exceptions (e.g., \texttt{sigill}, \texttt{sigbus}, \sigfpe, and \sigsegv). 
Openenclave only allows a small subset of signals (\texttt{sigill}, \texttt{sigbus}, \sigfpe, and \texttt{sigsegv}) to be forwarded to the enclave through its signal handling interface.

\subsection{Programming Languages}
\label{ssec:analysis-langs}

\begin{table}[]
  \caption{Compiler/Interpreter version.}
    \label{tab:compiler-version}
    \centering
\resizebox{0.5\columnwidth}{!}{%
\begin{tabular}{ll}
\toprule

Language & Version \\ \hline
C        & gcc (GCC) 13.2.1   \\
C++      & gcc (GCC) 13.2.1   \\
Java     & OpenJDK 17.0.10    \\
Python   & Python 3.11.6   \\ 
Go       & go 1.21.6     \\ 
Node.js   & v21.6.1   \\ 
Rust   & cargo 1.75.0   \\ 
Wasm   & wasmtime-cli 17.0.0      \\ 
Julia & julia version 1.10.2 \\
\bottomrule
\end{tabular}%
}
\end{table}

\begin{table}[]
  \caption{Implicit: Runtime converts signals into s/w exceptions. Explicit: Allow processes to register signal-specific handlers.}
    \label{tab:analysis-lang}
\resizebox{\columnwidth}{!}{%
\begin{tabular}{lll|lll}
\hline
Language & Implicit & \multicolumn{1}{c|}{Explicit} & Language & Implicit & \multicolumn{1}{c}{Explicit} \\ \hline
C        & \xmark   & \cmark                        & Go       & \xmark   & \cmark                       \\
C++      & \xmark   & \cmark                        & JS   & \xmark   & \cmark                       \\
Java     & \cmark   & \xmark                        & Rust     & \xmark   & \cmark                       \\
Python   & \xmark   & \cmark                        & Wasm     & \xmark   & \xmark                       \\ 
Julia  & \cmark   & \xmark                        &          &          &                              \\ \hline
\end{tabular}%
}
\end{table}

The SDKs and library OSes that we analyzed allow developers to execute programs written in different languages in the enclaves. 
While some of them provide support for specific languages (e.g., Teaclave is used to develop and run Rust programs), others support a wide range of languages (e.g., Scone supports executing programs written in C, C++, Java, Python, Rust, Go, JavaScript).\footnote{We analyze  server-side JS, specifically Node.js}
Therefore, we analyzed $9$ popular programming languages (see ~\cref{tab:compiler-version}) to check if they allow applications to register and execute signal handlers. 
To do this, we wrote programs in each of the languages to register signal handlers for all signals from $1$ to $31$. 
We report if the corresponding signal handlers were executed for each programming language in~\cref{tab:lang-extended-analysis}. 
The language and signal pairs that execute the registered signal handlers are of specific interest for \codename. 
An attacker can use these signals to compromise applications written in the corresponding language. 
For completeness, we also analyzed which signal handlers the programming language standard libraries or interpreters register by default i.e., when the program does not register any signal. 
We report our findings in ~\cref{tab:analysis-lang-sigs-default}. 
Our experiments show that signal number 10 (\sigusrone) and signal number 3 (\texttt{sigquit}) in Julia and Java respectively write debug logs to \texttt{stderr} and \texttt{stdout}. In \codename, \texttt{stdout} and \texttt{stderr} are not accessible to the attacker and are therefore not interesting. 
Further, Go starts profiling on signal 27 (\texttt{sigprof}), however this does not lead to any changes in the program's global state. 
Interestingly, NodeJS starts a debug server on signal 10 (\sigusrone). 
We exploit this behavior to demonstrate \codename on a NodeJS application in~\cref{ssec:nodejs}. 
Finally, we find that all other signal handlers that programming language runtimes register do not perform any computation that changes the program’s global state and simply crash the application. These handlers are therefore not interesting to \codename. 

\begin{figure}
    \centering
    \includegraphics[scale=0.7]{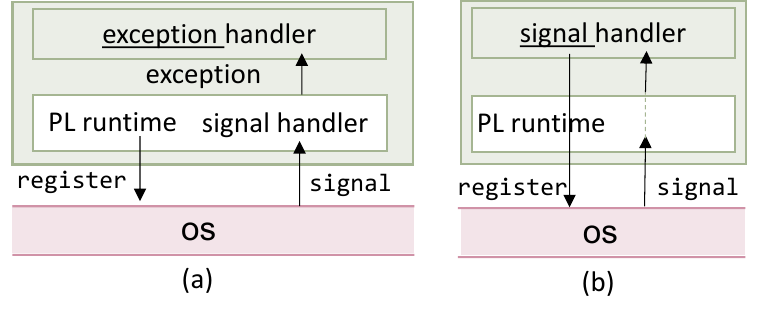}
    \caption{(a) Languages with implicit support for signals. The PL runtime registers signal handlers with the OS and converts signals from the OS to exceptions for the applications. (b) Languages with explicit support for signals. The PL runtime allows the applications to directly register handlers with the OS which are invoked when OS sends signals.}
    \label{fig:pl-analysis}
\end{figure}

\noindent{\bf Our findings.}
WebAssembly system interface (WASI), the standard interface definition for WebAssembly, does not yet support signals~\cite{wasi-signal-support}. 
So, programs compiled to Wasm binaries which use WASI cannot register or execute signal handlers.
For the other languages we broadly classify the signal support into $2$ categories: explicit and implicit support (\cref{fig:pl-analysis}). 
Programming languages that offer explicit signal support allow applications to register signal handlers for specific signals (e.g., using the \texttt{signal(signum,handler)} libc function) which are executed when the OS sends a signal to the enclave.
Therefore, \codename can asynchronously trigger these handlers when enclave programs  are written in these languages.

Languages which provide implicit support for signals (e.g., Java, Julia) register signal handlers directly with the OS. 
They do not provide any constructs for the programs to register signal-specific handlers. 
Instead, the language runtime converts signals into software exceptions. 
These software exceptions are caught and handled by the applications. 
For example, Java converts \sigfpe to \arithmeticexception and forwards it to the application. 
The application can execute custom handling for the exception in its \texttt{catch} block (e.g., Line 9 in \cref{fig:java-attack}).
Similarly, Julia converts \sigfpe to \divideerror which the application can catch and handle.
\codename can trigger these \texttt{catch} blocks to change the execution and data integrity of applications written in these programming languages. 

Go uses \texttt{sigprof} for internal CPU profiling and so allows programs to register handlers for all other signals except \texttt{sigprof}.
All other languages that provide explicit signal support, allow programs to register handlers for all signals between $1$-$31$ except $9$ and $19$ (i.e, \texttt{sigkill} and \texttt{sigstp}). 
Note that, this is a direct consequence of the kernel blocking all requests to register handlers for \texttt{sigkill} and \texttt{sigstp}. %
On the other hand, Java only converts {\tt sigfpe} to \arithmeticexception. 
Therefore, \codename can only use \sigfpe to compromise programs written in Java.  

In summary, of the $9$ languages that we analyzed, we found that $1$ does not support any signal handling, $2$ only supports implicit signal handling, and the $6$ others support explicit signal handling. 
Programs written in any of the $8$ languages that provide signal support may be vulnerable to \codename and should be reanalyzed in light of our findings.

\section{Case studies}
\label{sec:case-studies}
We confirmed our findings from~\cref{sec:sdks}-\cref{sec:analysis} using hand-coded enclaves. In this section, we first discuss end-to-end case studies and
explain how \codename can be used to compromise the security when executing in enclaves.
Then, we build a framework and demonstrate the feasibility of \codename{}'s signal injection architecture to show that it can be used to build exploits that require a large number of signal injections. 

\begin{figure}
    \centering
    \includegraphics[scale=0.55]{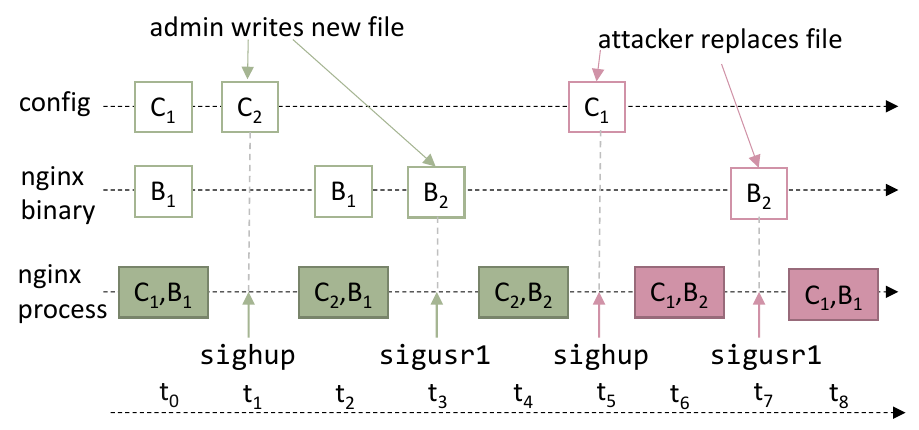}
    \caption{Maliciously injecting signals to \nginx to trigger insecure states (pink). C: Configuration, B: Binary.}
    \label{fig:example-nginx}
\end{figure}

\subsection{\nginx}
\label{ssec:nginx}

We surveyed widely used open-source webservers optimized for high uptime and found that many of them use signals to upgrade configuration without have to restart the serve. 
Specifically, httpd and Nginx use \sigusrone, and squid proxy server uses \sighup to upgrade the server's configuration. 
We choose to demonstrate \codename on Nginx as it is ported by library OSes (Gramine and Scone) to run in SGX enclaves. 

By default, \nginx allows a system administrator to upgrade its configuration and binaries using signals (\sighup and \sigusrone) without degrading the uptime of the server. 
For \codename, this gives the attacker the ability to change the configuration and binary of the server by injecting \sighup and \sigusrone. 
Gramine, Scone, and EnclaveOS port Nginx to run in SGX enclaves. 
Our analysis from~\cref{ssec:analysis-runtimes} shows that \codename cannot inject 
\sigusrone into enclaves running with Gramine. 
Further, the port of \nginx by EnclaveOS reduces the functionality of \nginx and does not allow administrators to refresh configuration files without restarting the server. 
Therefore, we demonstrate \codename on Nginx using Scone. 

\noindent{\bf Benign \nginx configuration and binary upgrade.}
Consider an Nginx webserver executing inside an enclave ($t_0$ in \cref{fig:example-nginx}). 
At time $t_1$, the \nginx administrator replaces the configuration file with an updated file and sends \sighup to the \nginx process. 
On receiving this, the \nginx process (at time $t_2$) reads the new configuration file and begins to use it. 
Similarly, the administrator upgrades the binary of the server by writing a new file and sending \sigusrone at time $t_3$ which is read and used by the server at $t_4$. 
Note that, in an enclave setting, the administrator encrypts the configuration and binary files before writing them to the OS accessible file-system. 
These files are only decrypted inside the enclave. 

\noindent{\bf Attacking \nginx with \codename.}
Because the administrator encrypts the configuration and binary files before writing them into the file-system, the OS cannot directly manipulate them to compromise \nginx.  
However, a malicious OS can record the encrypted blobs of configuration and binary files when the administrator writes them. 
The OS can also observe the new configuration and binary files when the administrator sends the signals to upgrade them in the enclave. 
Once the administrator has finished the configuration and binary upgrade, the OS uses \codename to manipulate the state of the \nginx server. 

Specifically, at $t_5$, the OS replaces the new configuration file with the old file it captured and sends malicious \sighup to the enclave. 
The enclave reads in this configuration and updates the \nginx process ($t_6$). 
This ensures that the \nginx process uses an old configuration with a new binary. 
Furthermore, the OS can also replace the new binary file with an old one that it captured and send \sigusrone ($t_7$). 
With this, the OS has successfully used \codename to restore the \nginx process to the state that it was in at time $t_0$ before the administrator performed the upgrades. 
This exploit will undo any performance and security improvements provided by the new binary and configuration and can be used by the OS to bring back old security bugs (e.g., forcing the enclave to use an \nginx version before v1.23.2  which patched critical issues~\cite{cve}).

\subsection{Node.js Server}
\label{ssec:nodejs}

Node.js is an open-source JavaScript runtime used for server-side scripting. 
By default, a Node.js server starts a debug web socket when it receives \sigusrone, even if the server was started without debugging enabled. 
To demonstrate \codename, we use Scone's Node.js port that executes an HTTPs server in an enclave.
When the server is up and running, the malicious OS sends \sigusrone to the Node.js server which opens a debug web socket. 
Using this web socket, the attacker can leak the server's memory and inject arbitrary code.
In our exploit, we use the web socket to leak the RSA private keys used for TLS by the server using $1$ injection of \sigusrone. 
Note that, a malicious network adversary cannot do this exploit (e.g., by sending a socket request for the debug socket to the \nodejs server).

\subsection{Multi-Normal Distribution}
\label{ssec:jsat}
We demonstrate \codename on a Java application, inspired by Heckler, to show how implicit signal support in programming languages make applications vulnerable~\cite{heckler-usenix}.
JSAT is a statistical analysis tool for machine learning applications in Java. 
JSAT implements a \texttt{MultiVariateNormal} class that can be used to create a Multivariate Gaussian distribution. 
The class implements a function that updates the mean and covariance of the distribution using the \texttt{setUsingData} function as shown in~\cref{lst:java-case-study}. If data added to the dataset causes an \texttt{ArithmeticException}, the \texttt{setUsingData} function discards the data and reverts the mean to the original value.
\begin{lstlisting}[language=Java, label={lst:java-case-study}, caption={JSAT mean and covariance.}]
public boolean setUsingData(List dataSet) {
 Vec origMean = this.mean;
 try { //can overflow
  Vec newMean = meanVector(dataSet);
  Matrix covarianc = covarianceMatrix(newMean,dataSet);
  this.mean = newMean; 
  setCovariance(covarianc);
 } catch(ArithmeticException ex) {
  this.mean = origMean;
 }
}\end{lstlisting}
As noted in~\cref{ssec:analysis-runtimes} and~\cref{ssec:analysis-langs}: (a) the Java runtime converts \sigfpe to ArithmeticException that can be caught and handled by Java applications, and (b) Gramine forwards \sigfpe to the enclaves. 
Therefore, we use \codename to trigger expressive application logic in JSAT executing in Gramine by injecting \sigfpe. 
We use one of the tests in JSAT to train a Learning Vector Quantization (LVQ) with Multivariate Gaussian distribution as a local classifier. 
This setup calls the \texttt{setUsingData} function during the training process of the classifier. 
We use \codename to inject \sigfpe every time the program executes lines $4-8$ in~\cref{lst:java-case-study}. 
In our attack, we inject \sigfpe $240$ times and drop the error rate of the classifier from nearly $0\%$ to $66\%$. 
Our attack adds an additional overhead of overhead of 3.4 seconds ($3.34\times$) to the training.
Therefore, \codename can be used to bias the classifier and consequently any inference that a user might execute on the model that it builds. 
We choose Java instead of Julia for our case study because library OSes support running Java applications in enclaves. 
However, if library OSes support Julia execution, a similar attack would be possible on Julia's text analysis framework~\cite{heckler-usenix}.

\subsection{Multi-Layer Perceptron}
\label{ssec:mlp}
To demonstrate the feasibility of \codename when an application requires a large number of signal injections we 
build a signal injection framework with sgx-step and Gramine.
To test the robustness of our framework, we identify an application that requires a large number of signal injections (on the order of $10^8$). 
To start with, we used an open-source implementation of a multi-layer perceptron written in C~\cite{mlp-git}. 
This library uses \texttt{tanh} as an activation function which is called over $10^8$ times during the training process.
Mathematically, $tanh(x)$ tends to $1$ when $x$ tends to $\infty$. 
Pragmatically, this occurs in the \texttt{tanh} function if the input overflows. 
To handle this, we introduce a signal handler to the tanh function as shown in~\cref{lst:tanh-function}.
We train the MLP model with 3 hidden layers and up to 6 units in each layer (tanh activation function in hidden layers, sigmoid in output layer). Our training consists of 2000 epochs and 1096 training samples from the Banknote data set~\cite{banknote-dataset}.
\begin{lstlisting}[language=C, label={lst:tanh-function}, caption={Tanh activation function for MLP.}]
jmp_buf buf;
void sigfpe_handler(int signum)
    longjmp(buf, 1);
void tanh(int input[...])
   ...
  for (i = 0; i < n; i++){ 
    if (sigsetjmp(buf, 1))
      output[i+1] = 1;               // on SIGFPE
    else 
      output[i+1] = tanh(input[i]);  // no overflow
\end{lstlisting}
Using \codename, if we inject \sigfpe when execution is in between lines $8$-$10$, it will trigger the signal handler and set \texttt{output[i+1]} to $1$ (Line 8). 
We inject \sigfpe into the \texttt{tanh} function $1,8636 \times 10^7$ times during the training of the neural network. 
Without \codename, the training achieves an accuracy of $97.09\%$. With \codename, the accuracy drops to $59.27\%$.
Further, \codename adds 970 seconds ($53.7\times$) overhead to the training process. 
This large overhead is primarily because of the asynchronous exits that sgx-step triggers and the signal handling in the enclave on every execution of the \texttt{tanh} function which the training invokes $\approx10^8$ times in total.

\section{Proof of Concept Exploits}

\begin{table}[]
\caption{Experiment Setup.}
\label{tab:setup}
\resizebox{\columnwidth}{!}{%
\begin{tabular}{@{}lcrrccccr@{}}
\toprule
Machine &
Intel CPU & Cores & RAM   & \begin{tabular}[c]{@{}c@{}}OS\\ (Ubuntu)\end{tabular}                                                      & Kernel & \begin{tabular}[c]{@{}l@{}}SGX\\ SDK\end{tabular}  & \begin{tabular}[c]{@{}l@{}}SGX\\ Driver\end{tabular} & \begin{tabular}[c]{@{}r@{}}Open\\ JDK\end{tabular} \\
\midrule

Laptop &\begin{tabular}[c]{@{}l@{}}Core i7-10875H\end{tabular} &   8    & 32 GB  & 20.04 & 5.10.2  & v2.16 & 2.11.0                                                & 17                                                  \\
Server & \begin{tabular}[c]{@{}l@{}}Xeon Gold 6346\end{tabular} & 32    & 378 GB & 20.04 & 5.19.0  & v2.16 & in-tree                                               & 17                                                  \\ \bottomrule
\end{tabular}%
}
\end{table}

We perform our experiments on machine setups as shown in~\cref{tab:setup}. 
We set up the laptop with sgx-step compatible configurations. 
The server has a newer kernel with an in-tree driver. 
Both machines have hardware support to store exit information in the SSA.

\noindent{\bf Sending Signals.} We send signals to enclaves with the \texttt{util-linux} program \texttt{kill}~\cite{kill}. The utility program is a wrapper around the \texttt{kill} syscall in Linux.
The \texttt{kill} syscall populates the \texttt{siginfo\_t} struct to indicate that the signal originated from user-space. 
Some of our experiments (e.g., injecting to Java applications) require the struct to indicate that the signal is the result of an integer division by zero error (\texttt{fpe\_intdiv}) or floating point overflow (\texttt{fpe\_fltovf}).
To do this, we implement a kernel module that correctly populates this information and sends it to the user-space process. (98 \loc).

\noindent{\bf \nginx.} 
Scone is closed-source and only supports running Nginx servers in the paid versions. Therefore, to perform our experiments we ported Nginx versions 1.22.1 and 1.24.0 to execute in a Scone v$5.8.0$ enclave~\cite{scone}. 
To do this, we remove some system calls  (e.g., \texttt{fcntl}) that Scone does not support. 
Further, we adapted Nginx to correctly propagate Scone configurations when spawning new processes (e.g., by invoking the \texttt{execve} system call).
Note that our modifications do not change the functional behavior of Nginx. 
We build Nginx with the \texttt{select} event method and a minimal configuration to serve HTTP websites.

\noindent{\bf Node.js.}
We run a Node.js v$10.14.1$ web server (28 \loc) with standard TLS libraries such as express and https using Scone's v$5.8.0$ publicly supported Node.js port and configuration~\cite{sconenodejs}. 
After sending \texttt{sigusr1} to Node.js, we use Chrome v$120.0.6099.224$ Developer tools to connect to the server, dump its memory, and extract the RSA private key used for TLS.

\noindent{\bf MLP and JSAT.}
To demonstrate \codename on the MLP training~(\cref{ssec:mlp}) we build a framework using sgx-step v$1.5.0$~\cite{sgx-step} and Gramine commit \texttt{211ec447e}~\cite{gramine-gh}.
First, as a preparatory step, we run the training of the model in Gramine in debug mode, to identify the instruction pages of the \texttt{tanh} function. 
To bias the training process, we must inject \sigfpe every time the \texttt{tanh} function is invoked. 
When the \texttt{tanh} function starts executing, sgx-step generates an asynchronous exit using timer interrupts and page faults. 
We should inject \sigfpe on this event but this is not straightforward.
On an asynchronous exit, control switches form the enclave to Gramine's uPAL. If we inject \sigfpe when uPAL is executing, Gramine will crash.
First, we change Gramine's uPAL (with 212 \loc)  to ignore our signals to avoid it from crashing.
Next, we use our framework to ensure that we inject \sigfpe only when the enclave has resumed executing the \texttt{tanh} function. 
For this, we use $2$ threads. The main thread executes the enclave and handles the asynchronous exit on each \texttt{tanh} invocation.
Then, a worker thread injects \sigfpe to the enclave after it is resumed.
Java applications require more profiling than ahead-of-time compiled languages (e.g., C).
For simplicity, we perform our experiments on JSAT by ensuring that our target function (\cref{lst:java-case-study}) waits for an \arithmeticexception on every invocation.

\section{Potential Defenses}
Current signal handling mechanisms render SDKs and library OSes vulnerable to \codename. We propose techniques to potentially address the root cause of these issues. Then, we consider orthogonal protection techniques that can be used to diminish \codename impact on a per-application basis.
\subsection{Detecting Fake Signals}
\noindent{\bf Hardware exceptions in SDKs.}
All the SDKs that we analyzed implement a hardware exception handling interface that checks the exit information before invoking application-registered exception handlers as shown in ~\cref{fig:sdk-exceptions}(a). 
Except for Intel SGX SDK, all of them introduce additional mechanisms to support enclave threads to send and receive signals.
This makes the SDKs vulnerable to \codename (\cref{fig:sdk-exceptions}(b)) and should be hardened. 
First, they should not accept signals from the OS that usually are a result of hardware exceptions (e.g., \sigfpe, \sigsegv) as these signals should be sent through the hardware exception handling interface and never through the signal handling interface for custom signals. 

\noindent{\bf Protecting inter-thread signals in SDKs.}
To prevent \codename, the SDKs should introduce a mechanism to check if a signal that a thread receives was in fact raised by an enclave thread. 
For example, the trusted runtime can set up a protected shared memory between the enclave threads. 
Then, to indicate that a signal was raised by an enclave thread, the trusted runtime can write the signal number and target threadID in the shared memory region and only then exit the enclave with an ocall.
Further, when the untrusted runtime enters the enclave with the signal (Step 5), the trusted runtime in the target thread can look up the shared memory region to check the legitimacy of the signal. 
This defense breaks functionality of applications that require signal injection from outside the enclave (e.g., upgrading \nginx configuration). 
If enclaves need signals from outside that are not raised for hardware exceptions for functionality, there is no mechanism to defend them against \codename. 

\noindent{\bf Hardware exceptions in library OSes.}
Gramine only supports signals that are raised because of hardware exceptions (c.f.~\cref{ssec:gramine}).
To stop \codename, Gramine's exception handling can be enhanced to use the exit information that the hardware stores in the SSA. 
Gramine has used this technique in a patch which stops \codename~\cite{gramine-patch}. 
Unlike Gramine, other library OSes support signals that are both raised because of hardware exceptions (e.g., \sigfpe) and explicitly by the processes (e.g., \sigusrone, \sighup).
Currently, library OSes do not distinguish between signals from hardware exceptions and signals that originate from within the application in the enclave. 
First, their signal handling logic should be separated for these two cases. 
Then, for the first case of hardware exceptions, the library OSes can filter illegitimate injections using the exit information stored by the hardware. 
Next, we discuss mechanisms to protect against injections of other signals that processes originate.

\noindent{\bf Interprocess-signals in library OSes.}
Our analysis in~\cref{sec:liboses} shows that library OSes that create distinct enclaves for application processes route the signal through the untrusted runtime and OS which makes them susceptible to \codename. 
Protecting this interface is challenging as enclave processes need a mechanism to check if the signal is from another trusted enclave process.  
SGX does not support setting up shared memory regions between multiple enclave processes. 
Therefore, a method similar to the one that we outlined for signals between threads is not straightforward to implement between enclave processes. 
To protect against \codename, the library OSes should implement a message passing framework to communicate the signal information to the target trusted processes which their library OS looks up to check the signal's legitimacy.

Occlum is vulnerable to \codename because it assumes a threat model where untrusted processes reside within the same enclave. 
While setting up a shared memory region to communicate the legitimacy of signals between processes in Occlum is easier, establishing the trust relationship between the processes is more challenging. 
To this end, Occlum should introduce a notion of attestation across process groups inside the same enclave.
Currently, all processes inside the enclave are mutually distrusting so this would be a fundamental architectural change to Occlum. 
With the trusted process groups notion, Occlum can use a message-passing mechanism to protect against \codename.
Finally, library OSes should not accept signals from untrusted software. 
If such signals that are not a result of hardware exceptions are required for functionality then there is no defense against \codename.

\subsection{ Limiting \codename Impact}
\codename relies on the fact that the signal is propagated to the enclave application by the runtime or library OS. 
Once the signal is sent to the application, it will execute the signal handler in the enclave which compromises the enclave's execution integrity. 
To amplify the effect of the signal handler, the attacker might need other capabilities.
For example, to compromise Nginx (\cref{ssec:nginx}), the attacker needs the ability to replace encrypted blobs of files to change the configuration or binary that the Nginx server uses. Similarly, for Node.js (\cref{ssec:nodejs}) the attacker needs to be able to connect to the port. 
If there are orthogonal protections like file-system rollback prevention, network firewalls, or dynamic attestation in place, these end-to-end attacks will be stopped. 
Of the library OSes we surveyed, only Gramine and Scone provide protections for filesystems and network. 
Gramine has filesystem protections to add trusted (i.e., hashed), and encrypted files to the enclave, but does not provide rollback protection. 
Scone paper does not discuss any port or protocol filtering in network protection for enclaves; when filesystem protection is enabled it ensures that the files are encrypted, rollback, and integrity protected. 
Next, we checked publicly available Scone versions---its network shields can be configured during enclave creation with a whitelist of allowed ports. 
However, this network shield is not activated by default allowing \codename to compromise Node.js. 
Thus, developers should consider the new attack surface introduced by \codename when configuring enclaves. 
Lastly, orthogonal protections cannot stop \codename attacks which exploit \sigfpe and compromise enclaves as these attacks directly affect enclave memory and do not need any external subsystem (e.g., network, files). 
Therefore, library OSes should implement comprehensive defenses for this new attack surface.

\subsection{Detecting \codename}
\codename injects signals into enclaves which cause asynchronous exits. 
One way to protect an enclave against \codename is to detect if an asynchronous exit was caused by fake signal injections. 
Intel introduced a hardware extension called AEX-notify to perform checks on reentry after an interrupt or exception~\cite{aex-notify-whitepaper}.
AEX-notify proposes a defense to prevent precise single-stepping of enclaves using the new hardware~\cite{aex-notify}. 
For this, they implement handlers for timer-interrupts that speed up the execution of the successive instructions. 
AEX-notify cannot mitigate \codename when the applications register handlers that only need to be invoked once (\cref{ssec:nodejs} and \cref{ssec:nginx}) to compromise the enclave.
Specifically, the OS can send the signal when the enclave has legitimately exited (e.g., timer interrupt, page-fault) without AEX-notify detecting it. 
This is because \codename will not cause any asynchronous exits, let alone 
require any single-stepping.

On the other hand, our attacks that need to send signals when the enclave application is executing a specific set of instructions can be harder, but not impossible, to perform in the presence of AEX-notify. 
However, this does not completely stop \codename. Note that, for \codename the attacker does not need to single-step the enclave. 
Instead, the attacker needs a mechanism to determine when the enclave is executing a set of instructions.
Further, if the size of this instruction set is larger than the number of instructions executed between timer interrupts, there will always be a legitimate enclave exit for the OS to then inject the signal. 
Then, when a genuine timer interrupt occurs in between these instructions, the attacker can inject the signal. 
For applications where this is not the case, the frequency of timer interrupts is controlled by the untrusted OS. Therefore, if needed, the attacker can tune the frequency to any value to send signals to enclaves when the enclave executes the target instruction set. 

\section{Related work}

\noindent{\bf Malicious Synchronous Interfaces.} 
Ports~\etal comprehensively analyze the threats to applications from an untrusted OS~\cite{overshadow-analysis}.
They perform this analysis in the context of Overshadow~\cite{Overshadow}, a framework that protects the confidentiality and integrity of applications in the presence of a malicious operating system and a trusted hypervisor. 
They note that because the untrusted OS manages the signals it can maliciously alter them. 
However, the analysis focuses on synchronous attacks from the OS that redirect signals or send bad return values. 
Iago builds on this, where the attacker {\em synchronously} manipulates the return values for system calls when the enclave makes {\em explicit} requests~\cite{checkoway2013iago}. 
These bad return values can be used to trick the enclave into performing unintended computations. 
Iago demonstrates how the limited sanitization in the system call interface between an enclave and OS can be used to synchronously alter enclave execution. 
This synchronous interface has been comprehensively studied and analyzed in the context of SGX enclaves~\cite{InkTag, coin, taleoftwoworlds, alder2023pandora}.

\noindent{\bf Malicious Asynchronous Interfaces.}
Previous works have demonstrated using {\em asynchronous} timer interrupts and page faults at arbitrary points of the enclave's execution~\cite{sgx-step}. The attacker manipulates the software to trigger these hardware events, which then trigger handlers that have {\em fixed} effects---timer interrupts and page faults cause an exit from the enclave, but do not execute any handlers in the enclave that change the program state.
In contrast, to the best of our knowledge, \codename is the first work on Intel SGX that injects signals and exceptions {\em asynchronously} at any point during the enclave's execution.
Depending on enclave-specific logic, such arbitrary signal injection can induce {\em varied effects} depending on the signal that is injected and the logic that the enclave has specified for it.
This allows \codename to execute such signal handlers at any point during enclave execution to bring about programmatic changes to the enclave's state.

\noindent{\bf Bugs in Enclave runtimes and Library OSes.}
Previous works have studied and demonstrated attacks on buggy enclave-OS interface implementations (e.g., Application Binary Interface, Application Programming Interface)~\cite{taleoftwoworlds, coin}. 
AsyncShock and Game of Threads exploit synchronization bugs in multi-threaded enclave applications by interrupting the threads at specific points in their execution~\cite{asyncshock} or using race-conditions~\cite{gameofthreads}. 
Along similar lines, SmashEx demonstrates attacks that use the lack of atomicity in signal handlers to compromise enclave execution~\cite{smashex2021}.
Unlike these previous works, \codename does not rely on bugs in the runtimes, library OSes, or signal handlers.

\noindent{\bf Bugs in Enclave applications.}
Prior works have used vulnerabilities in enclave application code to compromise it using code-reuse attacks~\cite{hackingindarkness}.
\codename does not rely on bugs in enclave application implementations. 
However, \codename can be used to bring back old bugs.
For example, by forcing \nginx to use old binaries, \codename brings back vulnerabilities in older \nginx versions. 

\noindent{\bf Detection Tools.}
Several works have used fuzzing~\cite{sgx-fuzz, FuzzSGX} and symbolic execution~\cite{alder2023pandora, TeeRex, coin, symgx} to detect vulnerabilities in enclaves. 
Notably, SEnFuzzer uses fuzzing to detect ocall and ecall interface bugs~\cite{SEnFuzzer}. 
We manually analyzed the runtimes and library OSes to find vulnerable runtimes, library OSes, and signal handlers that \codename can exploit. 
These detection tools can be used to further analyze applications with the intention of finding other effects of signal handling in enclaves. 
Lefeuvre~\etal, investigate the fragility of interfaces for software compartmentalization and build a fuzzer to detect interface vulnerabilities but do not consider signals~\cite{lefeuvre2022assessing}. Beyond Intel SGX, \codename's observations can be applied to other such avenues of research.

\noindent{\bf Side-channels.}
Prior works have used timer interrupts and page faults to leak side-channel information~\cite{van2017telling, xu2015controlled, he2018sgxlinger,  frontalattack, nemesis}.
Similarly, others have leveraged other side-channels to compromise enclaves~\cite{cacheattacks, Varys}. 
Sgx-step is a framework that enables attackers to precisely single-step enclaves~\cite{sgxstep-acsac}. 
While \codename is not a side-channel attack, side-channel information can be used to amplify the effects of \codename.

\noindent{\bf Kernel bugs due to interrupts and signals.}
Buggy implementation of signal handlers in the kernel can compromise the security of applications by inducing memory-corruption from race conditions in the handlers~\cite{owasp-signal-handler}.
ExpRace exploits kernel races by using interrupts in a non-TEE setting~\cite{ExpRace}. 
Prior works have analyzed the possibility of using buggy signal handlers and interrupt remapping to corrupt applications executing with trusted OSes~\cite{delivering-signals-for-fun-and-profit, msi-inj-attack}. 
Signal handlers have been known to have race condition bugs that can be exploited to compromise applications~\cite{owasp-signal-handler}. 
\codename does not assume any buggy or racy signal handlers.

\noindent{\bf Malicious Interrupts.}
There are three ways of delivering notifications to a user-level computation: 
interrupts, exceptions, and signals.
Since Intel TDX and AMD SEV offer a VM abstraction, the attacker can only use interrupts as a notification mechanism. Specifically, when the hypervisor delivers an interrupt, the guest OS can either handle it in its kernel with an interrupt handler or convert it to an exception or signal for the user process currently executing on the vCPU.  
A recent series of works dubbed Ahoi attacks, exploit CVMs provided by AMD-SEV and Intel TDX by injecting malicious interrupts from hypervisors~\cite{heckler-usenix, wesee-oakland}. In particular, Heckler injects interrupts which are then delivered as signals to the victim user-level program. 
\codename is inspired by Ahoi attacks, but investigates Intel SGX which is a user-level abstraction. 
\codename shows that a malicious OS can deliver interrupts, exceptions, and signals to an enclave.
When a CPU executing an enclave triggers a hardware interrupt (e.g., INT0 for divide by zero faults), Intel SGX hardware sets the SSA to indicate that the interrupt did indeed originate on the core. This way, the enclave software can check the SSA before executing the handler. 
Fortunately, our investigation shows that all current enclave SDKs (Intel SGX SDK, Open Enclave, Teaclave SGX-SDK) do indeed perform the interrupt authenticity check before executing the handler.
However, this covers only a fraction of the attack surface. 
The OS can still deliver software-based exceptions and signals directly to the enclave. Since these notifications are not based on hardware-events (e.g., executing an illegal instruction) triggered by the enclave, the CPU cannot provide any information about their authenticity.
Thus, \codename is an instance of Ahoi attack that uses exceptions and signals as the notification mechanism to compromise Intel SGX enclaves. 

\noindent{\bf \codename on Arm TEEs.}
Besides SGX, several interface attacks have been demonstrated on Arm TrustZone~\cite{boomerang, hpe}.
Further, prior works have used fuzzing to test TrustZone's trusted software~\cite{partemu}.
Unlike SGX, TrustZone does not expose signal interfaces to the malicious OS and filters interrupts from the untrusted OS to the trusted applications~\cite{TZOS}. 
Therefore \codename cannot be used to attack TrustZone.
Heckler shows that Arm CCA's protected VMs are not vulnerable to attacks using malicious interrupts because of Arm's interrupt architecture~\cite{heckler-usenix}. 
Specifically, Arm's interrupt architecture does not map interrupts to exceptions. 
Therefore, a malicious hypervisor cannot abuse interrupts to induce exceptions or trigger signal handlers in Arm CCA's VM. 
Furthermore, in Arm CCA's protected VM setting, the guest OS is trusted and the VM does not accept exceptions and signals from the untrusted hypervisor. 
Arm CCA is not vulnerable to \codename.

\section{Conclusion}
\codename 
demonstrates that a malicious OS can exploit Intel SGX enclaves by delivering malicious exceptions and signals to trick the enclave into executing handlers.
Our analysis of various runtimes and library OSes shows that they are vulnerable to \codename. 
Programming languages as well as native and ported enclave-bound programs that need exceptions and signal handling should consciously to choose between functionality and security.

\bibliographystyle{IEEEtranS}
\bibliography{paper}

% Generated by IEEEtranS.bst, version: 1.12 (2007/01/11)
\begin{thebibliography}{10}
\providecommand{\url}[1]{#1}
\csname url@samestyle\endcsname
\providecommand{\newblock}{\relax}
\providecommand{\bibinfo}[2]{#2}
\providecommand{\BIBentrySTDinterwordspacing}{\spaceskip=0pt\relax}
\providecommand{\BIBentryALTinterwordstretchfactor}{4}
\providecommand{\BIBentryALTinterwordspacing}{\spaceskip=\fontdimen2\font plus
\BIBentryALTinterwordstretchfactor\fontdimen3\font minus
  \fontdimen4\font\relax}
\providecommand{\BIBforeignlanguage}[2]{{%
\expandafter\ifx\csname l@#1\endcsname\relax
\typeout{** WARNING: IEEEtranS.bst: No hyphenation pattern has been}%
\typeout{** loaded for the language `#1'. Using the pattern for}%
\typeout{** the default language instead.}%
\else
\language=\csname l@#1\endcsname
\fi
#2}}
\providecommand{\BIBdecl}{\relax}
\BIBdecl

\bibitem{owasp-signal-handler}
``{Unsafe function call from a signal handler },''
  \url{https://owasp.org/www-community/vulnerabilities/Unsafe_function_call_from_a_signal_handler},
  {accessed 17.04.2024}.

\bibitem{gramine-patch}
``{[PAL/Linux-SGX] Cross-verify SW signals vs HW exceptions},''
  \url{https://github.com/gramineproject/gramine/commit/a390e33e16ed374a40de2344562a937f289be2e1},
  {accessed 22.04.2024}.

\bibitem{asylo}
``{Asylo Github},'' \url{https://github.com/google/asylo}, {accessed
  28.01.2024}.

\bibitem{aex-notify-whitepaper}
``{Asynchronous Enclave Exit Notify and the EDECCSSA User Leaf Function},''
  \url{https://www.intel.com/content/www/us/en/content-details/736463/white-paper-asynchronous-enclave-exit-notify-and-the-edeccssa-user-leaf-function.html},
  {accessed 28.01.2024}.

\bibitem{banknote-dataset}
``{Banknote Database},'' \url{https://banknotedb.com}, {accessed 28.01.2024}.

\bibitem{cve}
``{CVE-2022-41741},'' \url{https://nvd.nist.gov/vuln/detail/CVE-2022-41741},
  {accessed 28.01.2024}.

\bibitem{delivering-signals-for-fun-and-profit}
``{Delivering Signals for Fun and Profit},''
  \url{https://lcamtuf.coredump.cx/signals.txt}, {accessed 28.01.2024}.

\bibitem{edgelessrt}
``{Edgelessrt Github},'' \url{https://github.com/edgelesssys/edgelessrt},
  {accessed 28.01.2024}.

\bibitem{ego}
``{Ego Github},'' \url{https://github.com/edgelesssys/ego}, {accessed
  28.01.2024}.

\bibitem{enarx}
``{Enarx Github},'' \url{https://github.com/enarx/enarx}, {accessed
  28.01.2024}.

\bibitem{enclaveos}
``{Fortanix Runtime Encryption® Platform},''
  \url{https://resources.fortanix.com/hubfs/Fortanix\_RTE\_Platform\_Whitepaper.pdf},
  {accessed 28.01.2024}.

\bibitem{gramine-gh}
``{Gramine Github},''
  \url{https://github.com/gramineproject/gramine/tree/211ec447ee69f16139520fc3a17c561c36a00943},
  {accessed 28.01.2024}.

\bibitem{intel-sdk-sgx}
``{Intel SDK},'' \url{https://github.com/intel/linux-sgx}, {accessed
  28.01.2024}.

\bibitem{jwt-rfc}
``{JSON Web Token (JWT)},''
  \url{https://datatracker.ietf.org/doc/html/rfc7519}, {accessed 28.01.2024}.

\bibitem{kill}
``{kill(1) — Linux manual page},''
  \url{https://man7.org/linux/man-pages/man1/kill.1.html}, {accessed
  28.01.2024}.

\bibitem{mlp-git}
``{Multi Layer Perceptron in C},''
  \url{https://github.com/manoharmukku/multilayer-perceptron-in-c}, {accessed
  28.01.2024}.

\bibitem{mystikos}
``{MystikOS Github},'' \url{https://github.com/deislabs/mystikos}, {accessed
  28.01.2024}.

\bibitem{openenclave}
``{Openenclave Github},'' \url{https://github.com/openenclave/openenclave},
  {accessed 28.01.2024}.

\bibitem{posix-standard}
``{POSIX.1-2017},''
  \url{https://pubs.opengroup.org/onlinepubs/9699919799.2018edition/},
  {accessed 28.01.2024}.

\bibitem{rustedp}
``{Rust EDP Github},'' \url{https://github.com/fortanix/rust-sgx/tree/master},
  {accessed 28.01.2024}.

\bibitem{scone-gh}
``{Scone},'' \url{https://sconedocs.github.io}, {accessed 28.01.2024}.

\bibitem{sconenodejs}
``{Scone Nodejs},'' \url{https://sconedocs.github.io/Nodejs}, {accessed
  28.01.2024}.

\bibitem{teaclave}
``{Teaclave Github},''
  \url{https://github.com/apache/incubator-teaclave-sgx-sdk}, {accessed
  28.01.2024}.

\bibitem{wasi-signal-support}
``{WASI: signal handling},''
  \url{https://github.com/WebAssembly/WASI/issues/166}, {accessed 28.01.2024}.

\bibitem{alder2023pandora}
F.~Alder, L.-A. Daniel, D.~Oswald, F.~Piessens, and J.~Van~Bulck, ``Pandora:
  Principled symbolic validation of intel sgx enclave runtimes.''

\bibitem{TZOS}
ARM, ``{Learn the Architecture: TrustZone for AArch64},''
  \url{https://developer.arm.com/architectures/learn-the-architecture/trustzone-for-aarch64/trustzone-in-the-processor},
  2021.

\bibitem{scone}
\BIBentryALTinterwordspacing
S.~Arnautov, B.~Trach, F.~Gregor, T.~Knauth, A.~Martin, C.~Priebe, J.~Lind,
  D.~Muthukumaran, D.~O{\textquoteright}Keeffe, M.~L. Stillwell, D.~Goltzsche,
  D.~Eyers, R.~Kapitza, P.~Pietzuch, and C.~Fetzer, ``{SCONE}: Secure linux
  containers with intel {SGX},'' in \emph{12th USENIX Symposium on Operating
  Systems Design and Implementation (OSDI 16)}.\hskip 1em plus 0.5em minus
  0.4em\relax Savannah, GA: USENIX Association, Nov. 2016, pp. 689--703.
  [Online]. Available:
  \url{https://www.usenix.org/conference/osdi16/technical-sessions/presentation/arnautov}
\BIBentrySTDinterwordspacing

\bibitem{checkoway2013iago}
S.~Checkoway and H.~Shacham, ``Iago attacks: why the system call api is a bad
  untrusted rpc interface,'' \emph{ACM SIGARCH Computer Architecture News},
  vol.~41, no.~1, pp. 253--264, 2013.

\bibitem{Overshadow}
X.~Chen, T.~Garfinkel, E.~C. Lewis, P.~Subrahmanyam, C.~A. Waldspurger,
  D.~Boneh, J.~Dwoskin, and D.~R. Ports, ``Overshadow: A virtualization-based
  approach to retrofitting protection in commodity operating systems,''
  \emph{SIGOPS Oper. Syst. Rev.}, vol.~42, no.~2, p. 2–13, Mar. 2008.

\bibitem{TeeRex}
\BIBentryALTinterwordspacing
T.~Cloosters, M.~Rodler, and L.~Davi, ``{TeeRex}: Discovery and exploitation of
  memory corruption vulnerabilities in {SGX} enclaves,'' in \emph{29th USENIX
  Security Symposium (USENIX Security 20)}.\hskip 1em plus 0.5em minus
  0.4em\relax USENIX Association, Aug. 2020, pp. 841--858. [Online]. Available:
  \url{https://www.usenix.org/conference/usenixsecurity20/presentation/cloosters}
\BIBentrySTDinterwordspacing

\bibitem{sgx-fuzz}
\BIBentryALTinterwordspacing
T.~Cloosters, J.~Willbold, T.~Holz, and L.~Davi, ``{SGXFuzz}: Efficiently
  synthesizing nested structures for {SGX} enclave fuzzing,'' in \emph{31st
  USENIX Security Symposium (USENIX Security 22)}.\hskip 1em plus 0.5em minus
  0.4em\relax Boston, MA: USENIX Association, Aug. 2022, pp. 3147--3164.
  [Online]. Available:
  \url{https://www.usenix.org/conference/usenixsecurity22/presentation/cloosters}
\BIBentrySTDinterwordspacing

\bibitem{aex-notify}
\BIBentryALTinterwordspacing
S.~Constable, J.~V. Bulck, X.~Cheng, Y.~Xiao, C.~Xing, I.~Alexandrovich,
  T.~Kim, F.~Piessens, M.~Vij, and M.~Silberstein, ``{AEX-Notify}: Thwarting
  precise {Single-Stepping} attacks through interrupt awareness for intel {SGX}
  enclaves,'' in \emph{32nd USENIX Security Symposium (USENIX Security
  23)}.\hskip 1em plus 0.5em minus 0.4em\relax Anaheim, CA: USENIX Association,
  Aug. 2023, pp. 4051--4068. [Online]. Available:
  \url{https://www.usenix.org/conference/usenixsecurity23/presentation/constable}
\BIBentrySTDinterwordspacing

\bibitem{costan2016intel}
V.~Costan and S.~Devadas, ``Intel sgx explained.'' \emph{IACR Cryptology ePrint
  Archive}, vol. 2016, no. 086, pp. 1--118, 2016.

\bibitem{smashex2021}
J.~Cui, J.~Z. Yu, S.~Shinde, P.~Saxena, and Z.~Cai, ``Smashex: Smashing sgx
  enclaves using exceptions,'' in \emph{Proceedings of the 2021 ACM SIGSAC
  Conference on Computer and Communications Security}, ser. CCS '21, 2021.

\bibitem{gotee}
\BIBentryALTinterwordspacing
A.~Ghosn, J.~R. Larus, and E.~Bugnion, ``Secured routines: Language-based
  construction of trusted execution environments,'' in \emph{2019 USENIX Annual
  Technical Conference (USENIX ATC 19)}.\hskip 1em plus 0.5em minus 0.4em\relax
  Renton, WA: USENIX Association, Jul. 2019, pp. 571--586. [Online]. Available:
  \url{http://www.usenix.org/conference/atc19/presentation/ghosn}
\BIBentrySTDinterwordspacing

\bibitem{cacheattacks}
\BIBentryALTinterwordspacing
J.~G\"{o}tzfried, M.~Eckert, S.~Schinzel, and T.~M\"{u}ller, ``Cache attacks on
  intel sgx,'' in \emph{Proceedings of the 10th European Workshop on Systems
  Security}, ser. EuroSec'17.\hskip 1em plus 0.5em minus 0.4em\relax New York,
  NY, USA: Association for Computing Machinery, 2017. [Online]. Available:
  \url{https://doi.org/10.1145/3065913.3065915}
\BIBentrySTDinterwordspacing

\bibitem{partemu}
\BIBentryALTinterwordspacing
L.~Harrison, H.~Vijayakumar, R.~Padhye, K.~Sen, and M.~Grace, ``{PARTEMU}:
  Enabling dynamic analysis of {Real-World} {TrustZone} software using
  emulation,'' in \emph{29th USENIX Security Symposium (USENIX Security
  20)}.\hskip 1em plus 0.5em minus 0.4em\relax USENIX Association, Aug. 2020,
  pp. 789--806. [Online]. Available:
  \url{https://www.usenix.org/conference/usenixsecurity20/presentation/harrison}
\BIBentrySTDinterwordspacing

\bibitem{he2018sgxlinger}
W.~He, W.~Zhang, S.~Das, and Y.~Liu, ``Sgxlinger: A new side-channel attack
  vector based on interrupt latency against enclave execution,'' in \emph{2018
  IEEE 36th International Conference on Computer Design (ICCD)}.\hskip 1em plus
  0.5em minus 0.4em\relax IEEE, 2018, pp. 108--114.

\bibitem{InkTag}
O.~S. Hofmann, S.~Kim, A.~M. Dunn, M.~Z. Lee, and E.~Witchel, ``Inktag: Secure
  applications on an untrusted operating system,'' \emph{SIGPLAN Not.},
  vol.~48, no.~4, p. 265–278, Mar. 2013.

\bibitem{sgx}
Intel, ``Intel software guard extensions,''
  \url{https://software.intel.com/content/www/us/en/develop/topics/software-guard-extensions.html},
  {accessed 28.01.2024}.

\bibitem{boomerang}
\BIBentryALTinterwordspacing
P.~Jiang, Q.~Wang, J.~Cheng, C.~Wang, L.~Xu, X.~Wang, Y.~Wu, X.~Li, and K.~Ren,
  ``Boomerang: {Metadata-Private} messaging under hardware trust,'' in
  \emph{20th USENIX Symposium on Networked Systems Design and Implementation
  (NSDI 23)}.\hskip 1em plus 0.5em minus 0.4em\relax Boston, MA: USENIX
  Association, Apr. 2023, pp. 877--899. [Online]. Available:
  \url{https://www.usenix.org/conference/nsdi23/presentation/jiang}
\BIBentrySTDinterwordspacing

\bibitem{FuzzSGX}
A.~Khan, M.~Zou, K.~Kim, D.~Xu, A.~Bianchi, and D.~J. Tian, ``Fuzzing sgx
  enclaves via host program mutations,'' in \emph{2023 IEEE 8th European
  Symposium on Security and Privacy (EuroS\&P)}, 2023, pp. 472--488.

\bibitem{coin}
\BIBentryALTinterwordspacing
M.~R. Khandaker, Y.~Cheng, Z.~Wang, and T.~Wei, ``Coin attacks: On insecurity
  of enclave untrusted interfaces in sgx,'' in \emph{Proceedings of the
  Twenty-Fifth International Conference on Architectural Support for
  Programming Languages and Operating Systems}, ser. ASPLOS '20.\hskip 1em plus
  0.5em minus 0.4em\relax New York, NY, USA: Association for Computing
  Machinery, 2020, p. 971–985. [Online]. Available:
  \url{https://doi.org/10.1145/3373376.3378486}
\BIBentrySTDinterwordspacing

\bibitem{hackingindarkness}
\BIBentryALTinterwordspacing
J.~Lee, J.~Jang, Y.~Jang, N.~Kwak, Y.~Choi, C.~Choi, T.~Kim, M.~Peinado, and
  B.~B. Kang, ``Hacking in darkness: Return-oriented programming against secure
  enclaves,'' in \emph{26th USENIX Security Symposium (USENIX Security
  17)}.\hskip 1em plus 0.5em minus 0.4em\relax Vancouver, BC: USENIX
  Association, Aug. 2017, pp. 523--539. [Online]. Available:
  \url{https://www.usenix.org/conference/usenixsecurity17/technical-sessions/presentation/lee-jaehyuk}
\BIBentrySTDinterwordspacing

\bibitem{ExpRace}
\BIBentryALTinterwordspacing
Y.~Lee, C.~Min, and B.~Lee, ``{ExpRace}: Exploiting kernel races through
  raising interrupts,'' in \emph{30th USENIX Security Symposium (USENIX
  Security 21)}.\hskip 1em plus 0.5em minus 0.4em\relax USENIX Association,
  Aug. 2021, pp. 2363--2380. [Online]. Available:
  \url{https://www.usenix.org/conference/usenixsecurity21/presentation/lee-yoochan}
\BIBentrySTDinterwordspacing

\bibitem{lefeuvre2022assessing}
H.~Lefeuvre, V.-A. B{\u{a}}doiu, Y.~Chien, F.~Huici, N.~Dautenhahn, and
  P.~Olivier, ``Assessing the impact of interface vulnerabilities in
  compartmentalized software,'' \emph{arXiv preprint arXiv:2212.12904}, 2022.

\bibitem{Varys}
\BIBentryALTinterwordspacing
O.~Oleksenko, B.~Trach, R.~Krahn, M.~Silberstein, and C.~Fetzer, ``Varys:
  Protecting {SGX} enclaves from practical {Side-Channel} attacks,'' in
  \emph{2018 USENIX Annual Technical Conference (USENIX ATC 18)}.\hskip 1em
  plus 0.5em minus 0.4em\relax Boston, MA: USENIX Association, Jul. 2018, pp.
  227--240. [Online]. Available:
  \url{https://www.usenix.org/conference/atc18/presentation/oleksenko}
\BIBentrySTDinterwordspacing

\bibitem{overshadow-analysis}
D.~R.~K. Ports and T.~Garfinkel, ``Towards application security on untrusted
  operating systems,'' in \emph{Proceedings of the 3rd Conference on Hot Topics
  in Security}, ser. HOTSEC'08.\hskip 1em plus 0.5em minus 0.4em\relax USA:
  USENIX Association, 2008.

\bibitem{frontalattack}
\BIBentryALTinterwordspacing
I.~Puddu, M.~Schneider, M.~Haller, and S.~Capkun, ``Frontal attack: Leaking
  {Control-Flow} in {SGX} via the {CPU} frontend,'' in \emph{30th USENIX
  Security Symposium (USENIX Security 21)}.\hskip 1em plus 0.5em minus
  0.4em\relax USENIX Association, Aug. 2021, pp. 663--680. [Online]. Available:
  \url{https://www.usenix.org/conference/usenixsecurity21/presentation/puddu}
\BIBentrySTDinterwordspacing

\bibitem{gameofthreads}
\BIBentryALTinterwordspacing
J.~R. Sanchez~Vicarte, B.~Schreiber, R.~Paccagnella, and C.~W. Fletcher, ``Game
  of threads: Enabling asynchronous poisoning attacks,'' in \emph{Proceedings
  of the Twenty-Fifth International Conference on Architectural Support for
  Programming Languages and Operating Systems}, ser. ASPLOS '20.\hskip 1em plus
  0.5em minus 0.4em\relax New York, NY, USA: Association for Computing
  Machinery, 2020, p. 35–52. [Online]. Available:
  \url{https://doi.org/10.1145/3373376.3378462}
\BIBentrySTDinterwordspacing

\bibitem{wesee-oakland}
B.~Schlüter, S.~Sridhara, A.~Bertschi, and S.~Shinde, ``{WeSee: Using
  Malicious \#VC Interrupts to Break AMD SEV-SNP},'' in \emph{IEEE S\&P}, 2024.

\bibitem{heckler-usenix}
B.~Schlüter, S.~Sridhara, M.~Kuhne, A.~Bertschi, and S.~Shinde, ``{Heckler:
  Breaking Confidential VMs with Malicious Interrupts},'' in \emph{USENIX
  Security}, 2024.

\bibitem{occlum}
\BIBentryALTinterwordspacing
Y.~Shen, H.~Tian, Y.~Chen, K.~Chen, R.~Wang, Y.~Xu, Y.~Xia, and S.~Yan,
  ``Occlum: Secure and efficient multitasking inside a single enclave of intel
  sgx,'' in \emph{Proceedings of the Twenty-Fifth International Conference on
  Architectural Support for Programming Languages and Operating Systems}, ser.
  ASPLOS '20.\hskip 1em plus 0.5em minus 0.4em\relax New York, NY, USA:
  Association for Computing Machinery, 2020, p. 955–970. [Online]. Available:
  \url{https://doi.org/10.1145/3373376.3378469}
\BIBentrySTDinterwordspacing

\bibitem{occlum-paper}
------, ``Occlum: Secure and efficient multitasking inside a single enclave of
  intel sgx,'' in \emph{Proceedings of the Twenty-Fifth International
  Conference on Architectural Support for Programming Languages and Operating
  Systems}, 2020, pp. 955--970.

\bibitem{hpe}
\BIBentryALTinterwordspacing
D.~Suciu, S.~McLaughlin, L.~Simon, and R.~Sion, ``Horizontal privilege
  escalation in trusted applications,'' in \emph{29th USENIX Security Symposium
  (USENIX Security 20)}.\hskip 1em plus 0.5em minus 0.4em\relax USENIX
  Association, Aug. 2020. [Online]. Available:
  \url{https://www.usenix.org/conference/usenixsecurity20/presentation/suciu}
\BIBentrySTDinterwordspacing

\bibitem{taleoftwoworlds}
\BIBentryALTinterwordspacing
J.~Van~Bulck, D.~Oswald, E.~Marin, A.~Aldoseri, F.~D. Garcia, and F.~Piessens,
  ``A tale of two worlds: Assessing the vulnerability of enclave shielding
  runtimes,'' in \emph{Proceedings of the 2019 ACM SIGSAC Conference on
  Computer and Communications Security}, ser. CCS '19.\hskip 1em plus 0.5em
  minus 0.4em\relax New York, NY, USA: Association for Computing Machinery,
  2019, p. 1741–1758. [Online]. Available:
  \url{https://doi.org/10.1145/3319535.3363206}
\BIBentrySTDinterwordspacing

\bibitem{sgxstep-acsac}
J.~Van~Bulck and F.~Piessens, ``Sgx-step: An open-source framework for precise
  dissection and practical exploitation of intel sgx enclaves.''

\bibitem{sgx-step}
\BIBentryALTinterwordspacing
J.~Van~Bulck, F.~Piessens, and R.~Strackx, ``Sgx-step: A practical attack
  framework for precise enclave execution control,'' in \emph{Proceedings of
  the 2nd Workshop on System Software for Trusted Execution}, ser.
  SysTEX'17.\hskip 1em plus 0.5em minus 0.4em\relax New York, NY, USA:
  Association for Computing Machinery, 2017. [Online]. Available:
  \url{https://doi.org/10.1145/3152701.3152706}
\BIBentrySTDinterwordspacing

\bibitem{nemesis}
\BIBentryALTinterwordspacing
------, ``Nemesis: Studying microarchitectural timing leaks in rudimentary cpu
  interrupt logic,'' in \emph{Proceedings of the 2018 ACM SIGSAC Conference on
  Computer and Communications Security}, ser. CCS '18.\hskip 1em plus 0.5em
  minus 0.4em\relax New York, NY, USA: Association for Computing Machinery,
  2018, p. 178–195. [Online]. Available:
  \url{https://doi.org/10.1145/3243734.3243822}
\BIBentrySTDinterwordspacing

\bibitem{van2017telling}
J.~Van~Bulck, N.~Weichbrodt, R.~Kapitza, F.~Piessens, and R.~Strackx, ``Telling
  your secrets without page faults: Stealthy page $\{$Table-Based$\}$ attacks
  on enclaved execution,'' in \emph{26th USENIX Security Symposium (USENIX
  Security 17)}, 2017, pp. 1041--1056.

\bibitem{symgx}
\BIBentryALTinterwordspacing
Y.~Wang, Z.~Zhang, N.~He, Z.~Zhong, S.~Guo, Q.~Bao, D.~Li, Y.~Guo, and X.~Chen,
  ``Symgx: Detecting cross-boundary pointer vulnerabilities of sgx applications
  via static symbolic execution,'' in \emph{Proceedings of the 2023 ACM SIGSAC
  Conference on Computer and Communications Security}, ser. CCS '23.\hskip 1em
  plus 0.5em minus 0.4em\relax New York, NY, USA: Association for Computing
  Machinery, 2023, p. 2710–2724. [Online]. Available:
  \url{https://doi.org/10.1145/3576915.3623213}
\BIBentrySTDinterwordspacing

\bibitem{asyncshock}
N.~Weichbrodt, A.~Kurmus, P.~Pietzuch, and R.~Kapitza, ``Asyncshock: Exploiting
  synchronisation bugs in intel sgx enclaves,'' in \emph{Computer
  Security--ESORICS 2016: 21st European Symposium on Research in Computer
  Security, Heraklion, Greece, September 26-30, 2016, Proceedings, Part I
  21}.\hskip 1em plus 0.5em minus 0.4em\relax Springer, 2016, pp. 440--457.

\bibitem{msi-inj-attack}
R.~Wojtczuk and J.~Rutkowska, ``{Following the White Rabbit: Software attacks
  against Intel (R) VT-d technology},''
  \url{https://invisiblethingslab.com/resources/2011/Software\%20Attacks\%20on\%20Intel\%20VT-d.pdf},
  2011.

\bibitem{xu2015controlled}
Y.~Xu, W.~Cui, and M.~Peinado, ``Controlled-channel attacks: Deterministic side
  channels for untrusted operating systems,'' in \emph{IEEE S\&P}, 2015.

\bibitem{SEnFuzzer}
\BIBentryALTinterwordspacing
D.~Yu, J.~Wang, H.~Fang, Y.~Fang, and Y.~Zhang, ``Senfuzzer: Detecting sgx
  memory corruption via information feedback and tailored interface analysis,''
  in \emph{Proceedings of the 26th International Symposium on Research in
  Attacks, Intrusions and Defenses}, ser. RAID '23.\hskip 1em plus 0.5em minus
  0.4em\relax New York, NY, USA: Association for Computing Machinery, 2023, p.
  485–498. [Online]. Available: \url{https://doi.org/10.1145/3607199.3607215}
\BIBentrySTDinterwordspacing

\end{thebibliography}
\end{document}